\documentclass[aps,notitlepage,twocolumn,nofootinbib,superscriptaddress,pra,10pt]{revtex4-2}
\usepackage{graphicx}

\usepackage[english]{babel}
\usepackage{todonotes}
\usepackage{algorithm}
\usepackage{algpseudocode}
\usepackage{amsmath, amssymb}
\usepackage[utf8]{inputenc}
\usepackage{hyperref}

\usepackage{float}
\usepackage[caption=false,listofformat=subsimple,labelformat=simple]{subfig}

\usepackage{dsfont}

\usepackage{orcidlink}
\usepackage[version=4]{mhchem}
\usepackage[T1]{fontenc}
\usepackage{lmodern}
\usepackage{braket} 
\usepackage{color}
 \newcommand{\norm}[1]{\left\lVert#1\right\rVert}
\newcommand{\etal}{\textit{et~al.} }
\makeatletter
\def\maketitle{
\@author@finish
\title@column\titleblock@produce
\suppressfloats[t]}
\makeatother

\begin{document}

\title{Efficient optimization and conceptual barriers in \\
variational finite Projected Entangled-Pair States}

\author{Daniel Alcalde Puente  \orcidlink{0000-0002-3519-5931}}
\thanks{Both first authors have contributed equally.}
\affiliation{Institute for Theoretical Physics, University of Cologne, D-50937 K\"oln, Germany}
\affiliation{%
 Forschungszentrum J\"ulich  GmbH, Institute of Quantum Control, Peter Gr\"unberg Institut (PGI-8), 52425 J\"ulich, Germany
}%
\author{Erik Lennart Weerda
\orcidlink{0009-0006-8139-8648}}%
\thanks{Both first authors have contributed equally.}
\affiliation{Institute for Theoretical Physics, University of Cologne, D-50937 K\"oln, Germany}

\author{Konrad Schröder \orcidlink{0009-0007-4705-5147}}%
\affiliation{Institute for Theoretical Physics, University of Cologne, D-50937 K\"oln, Germany}
\affiliation{%
 Forschungszentrum J\"ulich  GmbH, Institute of Quantum Control, Peter Gr\"unberg Institut (PGI-8), 52425 J\"ulich, Germany
}%

\author{Matteo Rizzi \orcidlink{0000-0002-8283-1005}}%
\affiliation{Institute for Theoretical Physics, University of Cologne, D-50937 K\"oln, Germany}
\affiliation{%
 Forschungszentrum J\"ulich  GmbH, Institute of Quantum Control, Peter Gr\"unberg Institut (PGI-8), 52425 J\"ulich, Germany
}%

\date{\today}

\begin{abstract}
Projected entangled pair states (PEPS) on finite two-dimensional lattices are a natural ansatz for representing ground states of local many-body Hamiltonians, as they inherently satisfy the boundary law of entanglement entropy. 
In this paper, we propose the optimization of PEPS via an improved formulation of the time-dependent variational principle (TDVP), namely the  minimum-step
stochastic-reconfguration scheme recently introduced for neural quantum states.
We further discuss possible numerical issues that might arise in such a sampling-based approach. 
In this context, we investigate the entanglement properties of random PEPS and find an entanglement phase transition. 
We note that on one side of this transition, we can identify positive random tensors as product states.
To demonstrate the power of the framework described in this paper, we apply the PEPS to study the notoriously challenging chiral spin liquids.
Moreover, we exhibit our approach's capability to naturally handle long-range interactions by exploring the phase diagram of Rydberg atom arrays with long-range interactions.
We further provide parallelized easy-to-use code, allowing the straightforward application of our method to general Hamiltonians composed of local interaction terms.
\end{abstract}
\maketitle

\section{Introduction}
Understanding the ground state properties of interacting many-body systems remains a challenging topic in condensed matter physics.
Tensor network methods~\cite{ORUS2014practical, Cirac21RevModPhys, Rizzi19_anthology} have emerged as a particularly fruitful approach to address this problem.
These methods originate from the development of the \textit{density-matrix renormalization group}~\cite{White92_DMRG} and the subsequent analysis using insights from quantum information theory~\cite{SCHOLLWOCK2011_DMRG_MPS, Eisert2010RevModPhys, LAFLORENCIE20161}.
In two spatial dimensions, the most natural tensor network ansatz for ground state studies are the \textit{projected entangled-pair states} (PEPS)~\cite{verstraete2004PEPS}.
PEPS are particularly suitable as they inherently satisfy a boundary-law scaling of the entanglement entropy.

In recent years the translation-invariant infinite PEPS (iPEPS) have been the subject of substantial method development~\cite{Jordan08_iPEPS,Corboz14_iPEPS,Corboz16_variational,Vanderstraeten16_gradient,Liao19_AD_PEPS}, establishing iPEPS as a standard tool for the investigation of ground state properties in two dimensions and allowing for an increasingly wide scope of applicability~\cite{Liao17_iPEPS_appl, Niesen17_iPEPS_appl, Chen18_iPEPS_appl, Chung19_iPEPS_appl, Ponsioen19_iPEPS_appl, lee2020_iPEPS_appl, Gauthe20_iPEPS_appl, Hasik22_CSL, Peschke22_iPEPS_appl, Ponsioen23_iPEPS_appl, Ponsioen23_iPEPS_appl2, Weerda24_FQH, Hasik24_iPEPS_appl, schmoll2024_iPEPS_appl}.
On finite lattices, several PEPS studies with different algorithms have been performed~\cite{Murg07fPEPS,Murg09fPEPS,Lubasch_2014fPEPS,Lubasch2014fPEPS}, including some utilizing gradient-based approaches~\cite{Noack23_Hubbard_fPEPS,ORourke2023_Rydberg_entanglement, Gray2018}. However, no standard numerical approach has been established so far due to various numerical challenges.
In parallel to these efforts, a completely different approach to the PEPS on finite systems was developed based on the statistical evaluation of expectation values using sampling~\cite{Wang2011_sampling_TN}. 
Later, gradient-based optimization methods~\cite{Liu2017_gradient_fPEPS,liu2025accurate} have been used in conjunction with these sampling techniques. These have been applied successfully to paradigmatic models of frustrated magnetism~\cite{Liu2018_fPEPS_SL, Liu2021_accurate_fPEPS, LIU2022_fPEPS}. 
A particularly promising development for large-scale calculations using the sampling-based approach has been the development of a highly parallelizable method for directly generating samples from the PEPS state vector~\cite{Vieijra21_directsampling_PEPS}.

In this work, we investigate several aspects of sampling methods for finite PEPS. 
We emphasize that the most effective gradient-based optimization scheme, which relies on the \textit{time-dependent variational principle} (TDVP)~\cite{sorella2007weak, haegeman2011time, Carleo17_NQS, Vanderstraeten19TDVP}
can be implemented more efficiently by incorporating recent developments from the neural network literature~\cite{chen2024minSR}, where it is also referred to as Stochastic Reconfiguration.
Moreover, we demonstrate that despite using multiple approximations in numerical computations, the sampling-based PEPS approach can remain variational, as highlighted in \cite{liu2024tensor}. 
Consequently, it enables the determination of rigorous upper bounds for the ground state energy.
We demonstrate that the commonly employed, albeit formally less rigorous, approximation method yields estimates that closely align with the strictly variational results, yet it offers significantly enhanced computational efficiency.

Next, we examine the computational complexity associated with contracting single-layer tensor networks, a crucial step in the sampling-based finite PEPS framework.
Our analysis reveals several findings.
First, we demonstrate that random PEPS undergo an entanglement phase transition, coinciding with a complexity transition previously identified in random single-layer tensor networks~\cite{chen2024signproblem}. On one side of this entanglement transition we find the random positive PEPS, in essence, to be product states.
Further, the computational difficulty of contractions encountered in physically motivated scenarios is significantly lower than the worst-case scenarios identified in random tensor networks \cite{chen2024signproblem}.
We find that the entanglement present in the quantum state does not dictate the contractibility of its samples. 
Additionally, we identify a quantity that effectively serves as a predictor for the contraction complexity of physical PEPS samples.

Finally, we move to the applications of the finite PEPS for challenging physical situations.
We show that we can successfully apply the proposed finite PEPS framework to a gapped Hamiltonian hosting a chiral spin-liquid as its ground state, a situation that has been notoriously difficult for PEPS.
Further, we point out that the finite PEPS sampling framework can treat very conveniently certain long-range interactions, which can be extremely challenging in alternative approaches. 
We utilize this to study the phases of a model with long-range interacting Rydberg-atom arrays and point out further possible applications.\\

The paper is organized as follows: 
In Sec.~\ref{sec:finite PEPS and sampling}, we present the sampling framework for Projected Entangled Pair States (PEPS) and introduce an improved approach to implementing the time-dependent variational principle (TDVP) within this sampling framework. 
In Sec.~\ref{sec:methods}, we detail the numerical methods employed and address aspects related to variational upper bounds and the approximations involved in PEPS calculations. 
Sec.~\ref{sec:contraction complexity issues in sampling PEPS} is dedicated to analyzing the complexity associated with tensor network contractions in sampling-based PEPS computations. 
Finally, in Sec.~\ref{sec:applications}, we apply the developed finite PEPS methods to several models, including chiral spin liquids and Rydberg atom arrays.

\section{Finite PEPS and Sampling}

\label{sec:finite PEPS and sampling}
Projected Entangled Pair States (PEPS) provide a powerful ansatz for studying quantum many-body systems on finite lattices of size $L_x \times L_y$. 
As they satisfy the boundary law of entanglement entropy in two dimensions, they are well-suited for representing ground and low-energy states of local Hamiltonians.

The fundamental building blocks of finite PEPS are local tensors, which are defined on each lattice site
\begin{equation} 
T[x,y] = \begin{gathered}
        \includegraphics[height=2.6cm]{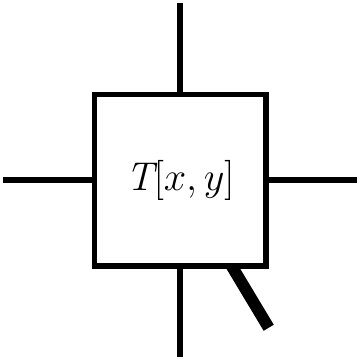}
\end{gathered}\;.\end{equation}
Every local tensor has a single \textit{physical} index (bold line), labeling a basis of its local physical Hilbert space of dimension $d$, as well as a set of \textit{virtual} indices (horizontal and vertical lines) of dimension $D$ that connect to the local tensors on neighboring sites. 
Within the PEPS ansatz, we express the coefficient tensor of the many-body state as the contraction of all virtual indices, illustrated as connected legs, of the local tensors
\begin{equation}
\label{PsiS}
\begin{split}
    \ket{\Psi} = \sum_{\{\textbf{S}\}} &\Psi(\textbf{S}) \ket{\textbf{S}}, \quad \ket{\textbf{S}}=\ket{s_{(1,1)}\dots s_{(L_x,L_y)}}\\
    \Psi(\textbf{S}) &= \begin{gathered}
        \includegraphics[height=3.8cm]{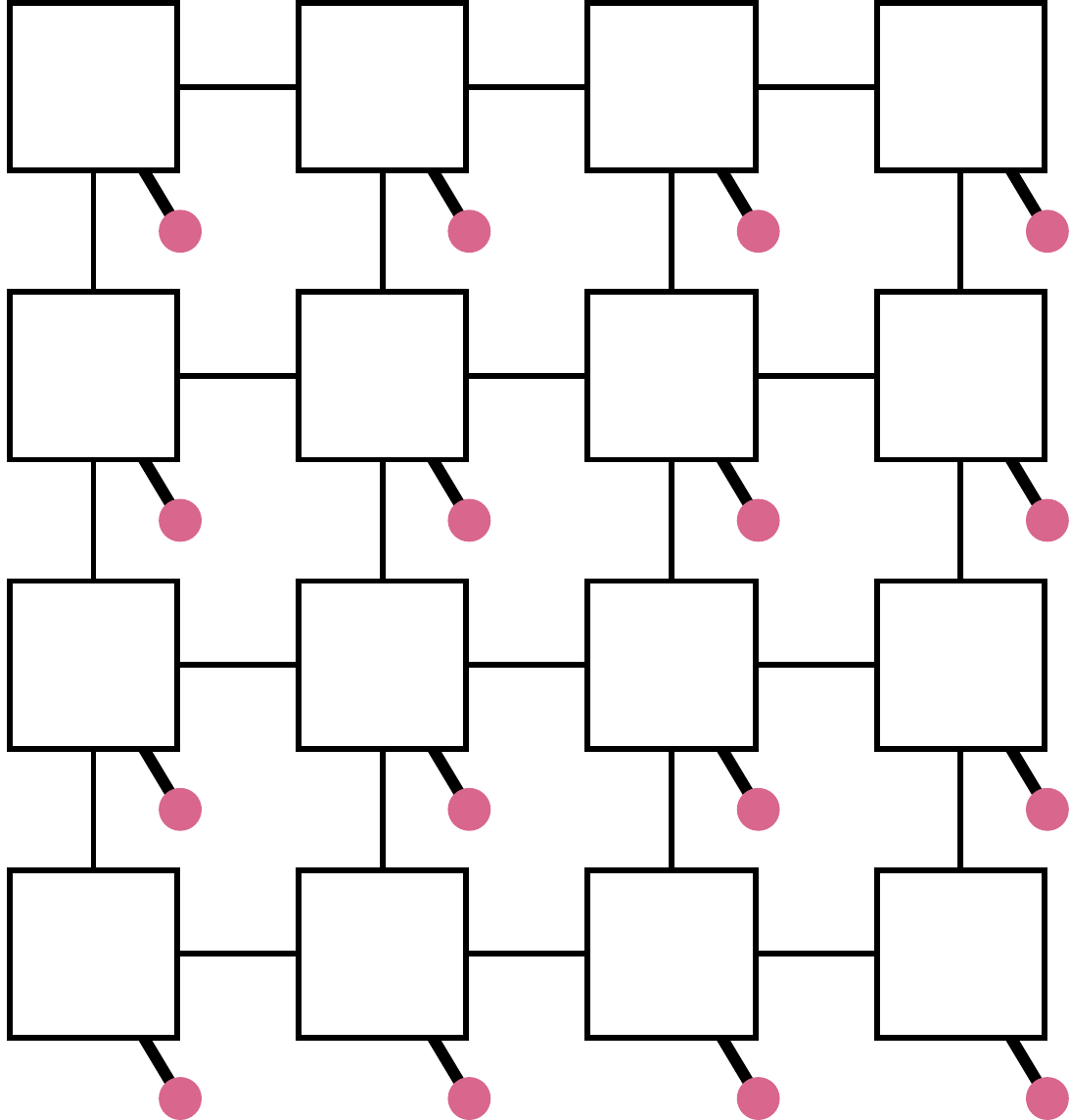}
    \end{gathered}
\end{split}
\end{equation}
Here, $\textbf{S}$ denotes a single many-body configuration, and the red dots signify that we fixed the index at position $(x,y)$ to a certain value $s_{(x,y)}$.

To calculate expectation values with finite PEPS, one in principle has to perform a contraction of a double-layer tensor network, in which one layer corresponds to the \textit{bra}- and the other to the \textit{ket}-vector. However, the exact contraction of a two-dimensional double-layer tensor network is computationally hard, such that schemes to perform the contraction approximately have to be used in practice, see e.g.~\cite{Lubasch_2014fPEPS}, which can still be challenging.

One alternative to this has been pioneered by Wang \etal~\cite{Wang2011_sampling_TN}, who demonstrated that one can use sampling techniques to statistically approximate expectation values of finite PEPS. 
In this framework, only the contraction of single-layer tensor networks is necessary, which can result in substantially cheaper numerical calculations.
A very similar sampling-based approach is also employed in methods using the neural quantum state ansatzes~\cite{Carleo17_NQS}.
To optimally make use of the statistical approach for ground-state calculations, we only need to be able to perform a few numerical operations. 
Firstly, in order to evaluate expectation values 
 \begin{equation}
 \begin{split}
     \label{eq:local_estimator}
     \braket{\hat{O}} = \frac{\braket{\Psi|\hat{O}|\Psi}}{\braket{\Psi|\Psi}} &= \sum_{\textbf{S}} \underbrace{\frac{|\Psi(\textbf{S})|^2}{\braket{\Psi|\Psi}}}_{p_{\Psi}(\textbf{S})} \cdot 
     \underbrace{\frac{\braket{\textbf{S}|\hat{O}|\Psi}}{\Psi(\textbf{S})}}_{O_\text{loc}(\textbf{S})}\\
    \bra{\textbf{S}}\hat{O}\ket{\Psi} &= \sum_{\textbf{S'}}O_{\textbf{S'}\textbf{S}}\Psi(\textbf{S'}),
 \end{split}
 \end{equation}
we need to be able to generate samples \textbf{S} from the state-vector according to $p_{\Psi}(\textbf{S})$, and further be able to calculate the coefficient $\Psi(\textbf{S})$ of the many-body state vector for a given sample.
Secondly, to optimize the ansatz (finite PEPS or neural quantum state) one needs the ability to calculate the gradient $\frac{\partial \Psi_{\theta}(\mathbf{S})}{\partial \theta}$ of each state vector coefficient w.r.t. its underlying variational parameters, which we label as $\theta$ and suppress in the expressions in general.
For the PEPS these variational parameters are the entries in the local tensors, which results, e.g.,  in  
\begin{equation}
   \frac{\partial \Psi_{\theta}(\mathbf{S})}{\partial T[3,2]} = \begin{gathered}
        \includegraphics[height=3.8cm]{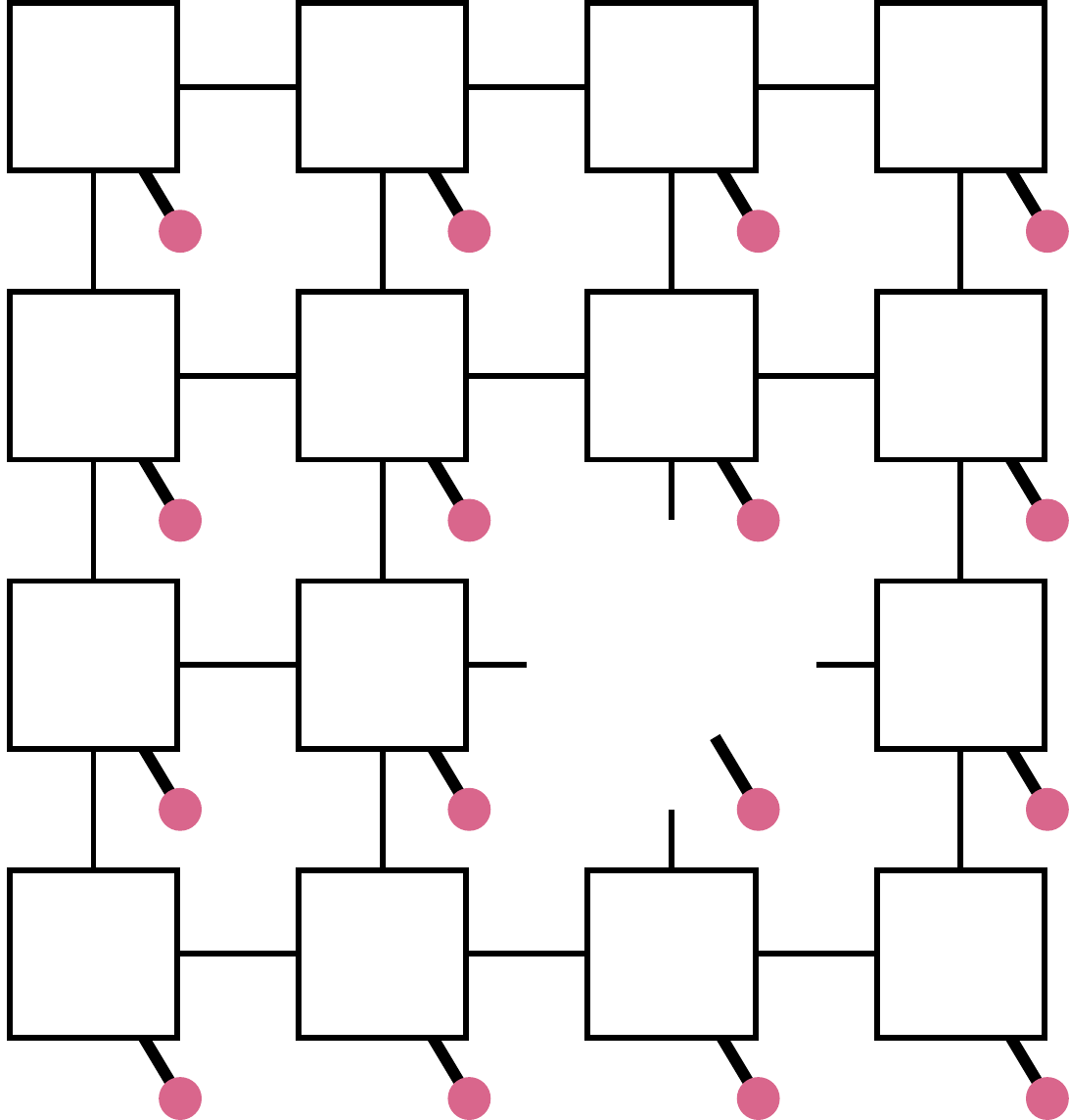}
    \end{gathered}\;.
\end{equation}
In Sec.~\ref{sec:methods} we discuss the numerical techniques to perform these necessary tasks and highlight in particular that, for local lattice models, the calculation of (energy) expectation values can be done in a distinctly more efficient way as compared to e.g. neural quantum states. 

\subsection{Reformulation of the imaginary-time Schrödinger equation for sampling techniques}
\label{sec:imaginary_schrodinger_eq}

To optimize the PEPS, a gradient-based scheme is used. Specifically, the time-dependent variational principle (TDVP)~\cite{sorella2007weak, haegeman2011time, Carleo17_NQS, Vanderstraeten19TDVP} is employed. It governs the evolution of parameters $\theta$ in a variational wave function $\ket{\Psi(\theta)}$ to ensure that the state vector evolves according to the Schrödinger equation as close as possible. Here, we are specifically interested in the use of the TDVP in the sampling-based approach for the evolution according to the imaginary-time Schrödinger equation. 
We note that, while we discuss the use for ground state search here, the same approach could be in principle used to simulate real-time evolution. 
We comment on related difficulties in App.~\ref{app:real_time_evo}.

We derive the equations for the evolution of the parameters $\theta$ by simple manipulation of the imaginary-time Schrödinger equation
\begin{align}
    \ket{\dot{\Psi}(\theta)} &= -H \ket{\Psi(\theta)} \label{eq:Schrodinger}\\
    \frac{\partial \ket{\Psi(\theta)}}{\partial \theta} \dot{\theta} &= -H \ket{\Psi(\theta)} \\
    \sum_i \underbrace{\frac{\partial \braket{\textbf{S}|\Psi(\theta)}}{\partial \theta_i}}_{\tilde{O}_{\textbf{S},i}} \dot{\theta_i} &= -\underbrace{\braket{\textbf{S}|H|\Psi}}_{\tilde{E}^\text{loc}_\textbf{S}} \quad \forall \, \mathbf{S}\\
    \tilde{O} \, \dot{\theta} &= \tilde{E}^\text{loc}. \label{eq:tdvp}
\end{align}
Note that if we write Eq.~\eqref{eq:Schrodinger} with an explicit normalization for $\ket{\Psi}$ we obtain an additional term.

If the number of parameters is smaller than the dimension of the Hilbert space the above equations will in general not have an exact solution. Instead, we are looking for the best approximation of a solution in the parameter space
\begin{align}
\min_{\dot{\theta}}& \sum_\mathbf{S} \norm{\sum_i\tilde{O}_{\textbf{S},i} \dot{\theta_i} - \tilde{E}^\text{loc}_\textbf{S}}^2\;.
\end{align}
Instead of finding the best approximation on the entire Hilbert space, we use Monte Carlo sampling to find a solution on the relevant, sampled part of the Hilbert space whose dimension is the number of samples $N_s$
\begin{align}
\min_{\dot{\theta}}\sum_\textbf{S} p_\Psi(\textbf{S}) \norm{\sum_i\underbrace{\frac{\tilde{O}_{\textbf{S},i}}{\Psi(\textbf{S})}}_{=:O_{\textbf{S},i}} \dot{\theta_i} - \underbrace{\frac{\tilde{E}^\text{loc}_\textbf{S}}{\Psi(\textbf{S})}}_{=:E^\text{loc}_\textbf{S}}}^2\;.
\end{align}
Conventionally this is solved by defining a matrix $G$ which is approximated on the samples drawn
\begin{align}
    \dot{\theta} &= (\underbrace{O^\dag O}_{=:G})^{-1} O^\dag E_\text{loc} \label{eq:tdvp_slow}\\
    G_{i',i} &= \sum_\textbf{S} p_{\Psi}(\textbf{S}) O^*_{\textbf{S},i'} O_{\textbf{S},i}
\end{align}
Unfortunately, the matrix $G$ has dimensions $N_p \times N_p$, where $N_p$ denotes the number of parameters in the ansatz. For parameter-intensive ansätze such as finite PEPS, where $N_p = L_x L_y D^4 d$, storing $G$ becomes infeasible. Consequently, slow iterative solvers are typically employed. Importantly, the rank of the sampled $G$ matrix is at most the number of samples, $N_s$.

\begin{figure}[t]
    \subfloat[]{\includegraphics[width=0.4\textwidth]{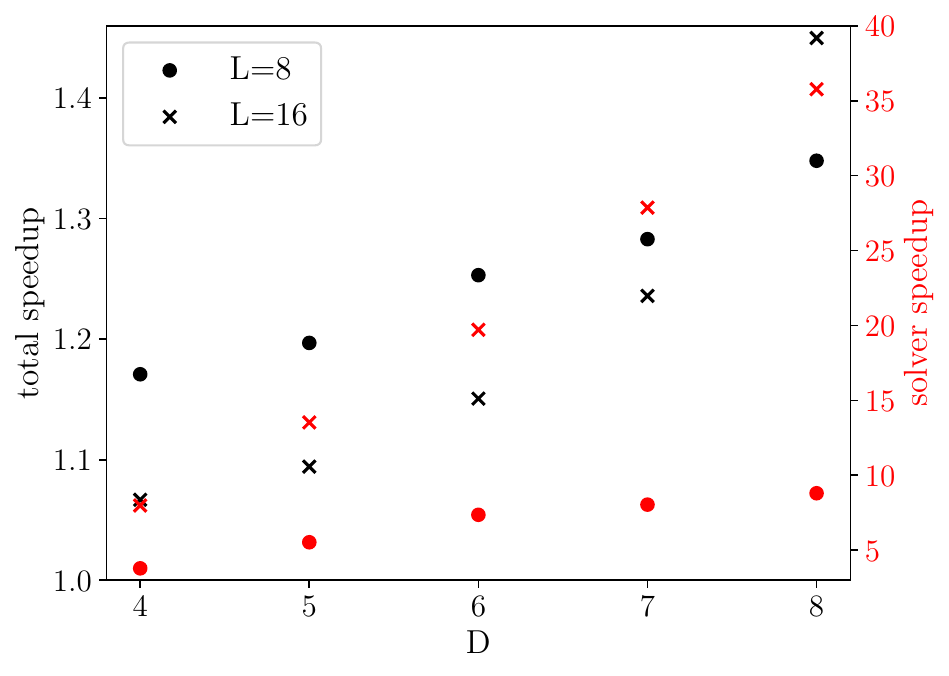}}
    \caption{Speedup of employing minSR vs a Krylov-based iterative solver for different bond dimensions for system sizes $L=8,16$ with $N_s=1000$ during the optimization. (black) Is the total time quotient $\frac{t_\text{total}[\text{krylov}]}{t_\text{total}[\text{minSR}]}$. (red) The quotient time for the solvers $\frac{t_\text{solver}[\text{krylov}]}{t_\text{solver}[\text{minSR}]}$}.
    \label{fig:timings_minsr}
\end{figure}

Recently, in the context of neural quantum states, an alternative method referred to as minimum-step
stochastic-reconfguration (minSR) was
proposed by Chen and Heyl~\cite{chen2024minSR}. It addresses the issue that arises for ansatzes with large numbers of parameters.
Their method works by leveraging the low rank of $G$ in the scenario where there are more parameters than samples
\begin{align}
    \dot{\theta} =& O^\dag (\underbrace{O O^\dag}_T)^{-1}  E_\text{loc} \label{eq:tdvp_fast}\\
    T_{\textbf{S}\textbf{S'}} =& \sum_i O_{\textbf{S},i} O^*_{\textbf{S'},i}.
\end{align}
Here, the matrix $T$ is only considered on the sampled subspace of the Hilbert space and hence has dimensions $N_s \times N_s$. 
Thus, it is significantly more manageable to store it and compute its pseudoinverse. 
In Fig.~\ref{fig:timings_minsr} we show that, 
when we apply this technique to finite PEPS, 
it substantially increases the efficiency of solving the linear system~\eqref{eq:tdvp} for $\dot{\theta}$, which is necessary in every optimization step.
The fraction of the total compute time of the ground state search with finite PEPS varies substantially for different use cases, but to get some idea of the speedup our code was timed for the $J_1$-$J_2$ model using both methods.

We observe up to a $40\%$ speedup in total compute time, and a $39$ times faster solving of the TDVP equation for $L=16$ PEPS. With higher bond dimensions and large system sizes yielding the most gains. Note that these percentages were computed by comparing a Krylov method to solve Eq.~\eqref{eq:tdvp_slow} and an eigenvalue decomposition method to solve Eq.~\eqref{eq:tdvp_fast}.

%


\section{Methods}
\label{sec:methods}
When employing the PEPS on a finite lattice in the sampling-based framework, we use several numerical tools, which are summarized in this section. We further comment on the approximations made in the sampling-based PEPS scheme by comparing energies to true variational upper bounds to the ground state energy, which can be obtained at a higher cost.

\subsection{On the boundary MPS method}
\label{sec:boundaryMPS}
In order to contract finite two-dimensional tensor networks, we use the boundary-MPS method~\cite{verstraete2004PEPS, Lubasch_2014fPEPS}.
This method can be applied to single-layer tensor networks, as they appear in Eq.~\eqref{eq:local_estimator}, as well as to double-layer network contraction, which is more costly, as we discuss below. 
We label the local tensors on a finite square lattice of size $L_x \times L_y$ by their positions in the lattice $T[x,y]$ and assume that they have an identical bond dimension $D$ for each of their virtual indices. 
The entries in the local tensors are the parameters of the PEPS ansatz $\theta=\{T[1,1],...,T[L_x, L_y]\}$.
The central objects of this contraction algorithm are the boundary-MPS environments, which approximate the contraction of a set of rows of the network starting from a boundary. 
We will, for completness, discuss here an example of an approximate contraction using boundary-MPS for the uppermost rows.
The boundary-MPS environment can be recursively defined so that it approximates the contraction of the $i$ uppermost rows of the tensor network as an MPS of bond dimension $D_c$:
\begin{equation}
\begin{split}
     E^{\text{u}}[i] :=& \;\begin{gathered}
        \includegraphics[height=1.85cm]{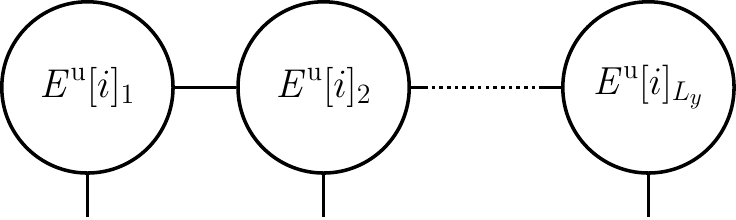}
        \end{gathered}\\[10pt]
    E^{\text{u}}[i]\approx& \;\begin{gathered}
        \includegraphics[height=3.79cm]{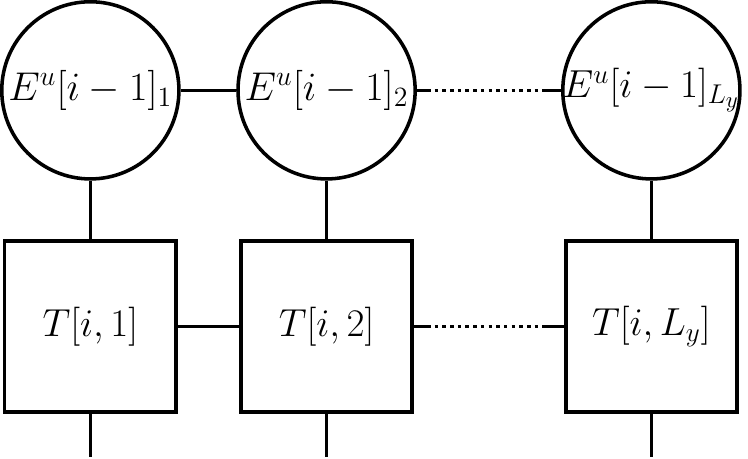}
    \end{gathered}\;.
\end{split}
\end{equation}
In the second row, the approximation indicates that we employ an MPS–MPO multiplication to truncate the bond dimension of the resulting MPS from \(D_c D\) back to \(D_c\). 
For this task, several algorithms have been developed in the past~\cite{Stoudenmire_2010_METTS, McCulloch_2007_DMRG_MPSMPO}. 
Here, we employ the density matrix algorithm~\cite{McCulloch_2007_DMRG_MPSMPO}.
For the initial boundary MPS on the top, the uppermost row of local tensors is used.
Of course, a boundary-MPS can be defined completely analogously for the approximate contraction for the lowermost rows up to row $j$, which we refer to as $E^l[j]$. 
These boundary-MPS can be combined to approximate the contraction of the network from both sides
\begin{equation}
\begin{split}
 \text{tTr}\biggl[\prod_{x,y} T[x,y]  \biggr] \approx 
 E^u[i] \cdot E^l[i+1], \; \forall \; i \in [1,L_x-1].
\end{split}
\label{eq:boundary_MPS_approx}
\end{equation}
where tTr is the tensor trace over all indices not shown explicitly.
As mentioned above, the boundary-MPS algorithm can be used both for single- as well as for double-layer tensor networks with different computational costs.
Specifically, for the application in the double layer case the cost scales as $\mathcal{O}(D_c^3 D^4) + \mathcal{O}(dD_c^2 D^6)$~\cite{Lubasch_2014fPEPS}, while for a single layer the dominant scaling is $\mathcal{O}(D_c^3 D^3) + \mathcal{O}(D_c^2 D^4)$. 
Let us note that the bond dimension of the boundary-MPS $D_c$ used in practical calculations should be increased whenever more accurate results are needed. Specifically, one can decide on $D_c$ dynamically by setting a fixed value for the maximal truncated weight, cf. App.~\ref{app:true_variational} for an analysis of the impact of different choices of the maximal truncated weight. Due to this analysis, in the following, a maximal truncated weight of $10^{-4}$ was used.
We refer to Sec.~\ref{sec:contraction complexity issues in sampling PEPS} for a discussion of the possible problems and usability of the boundary-MPS algorithm for different single-layer situations.
Lastly, we mention that the boundary-MPS are also used for the calculations of gradients, which are necessary for optimization. 

\subsection{On reusing environments and the convenient treatment of long-range interactions}
\label{sec:reusing_envs}

When using the boundary-MPS algorithm to compute the components of Eq.~\eqref{eq:local_estimator}, it is often possible to reduce the computational cost by avoiding the repeated calculation of boundary-MPS environments. For local operators that act non-trivially only on a limited number of sites, the sum in
\begin{equation}
    O_\text{loc}(\textbf{S})= \sum_{\textbf{S'}}O_{\textbf{S'}\textbf{S}}\frac{\Psi(\textbf{S'})}{\Psi(\textbf{S})}, \label{eq:local_obs}
\end{equation}
is restricted to configurations $\textbf{S'}$ that differ from $\textbf{S}$ only by a few modified elements within the support of $\hat{O}$. Consequently, a significant computational advantage of PEPS, compared to neural quantum ansätze, lies in the ability to reuse precomputed environments for $\Psi(\textbf{S})$ on lattice sites outside the support of $\hat{O}$.

If the operator $\hat{O}$ acts non-locally, reusing boundary-MPS environments becomes less effective. However, if $\hat{O}$ is diagonal in the computational basis, $O_\text{loc}(\mathbf{S})$ in Eq.~\eqref{eq:local_obs} becomes independent of $\Psi(\textbf{S})$. In this case, no additional computations are required. This property is particularly advantageous in Sec.~\ref{sec:Rydbergs}, where it facilitates the treatment of long-range interactions in Rydberg atom arrays.

This approach can also be applied when the operator is not entirely diagonal. For instance, consider a Hamiltonian consisting of long-range Heisenberg interactions. The terms of this Hamiltonian can be divided into three sets: one diagonal in the $X$-basis, one in the $Y$-basis, and one in the $Z$-basis.
To evaluate the expectation values of the operators in each set, one can transform the PEPS into the corresponding diagonal basis by a global spin rotation. In a diagonal basis, the calculations become straightforward. Note, that one has to sample independently for each of the transformed PEPS.
Hence, this procedure becomes less effective if the Hamiltonian consists of more sets of operators that do not share a common diagonalizing basis.

\subsection{On Numerical Stability}
The method used to contract single-layer samples of finite PEPS, $\Psi(\textbf{S})$, involves numerous MPS-MPO multiplications. 
To ensure numerical stability during these operations, each tensor $(E^{u/l}[i])_j$ of the environment is normalized. 
This is achieved by dividing the tensor by a factor $f_{ij}$ such that its largest entry equals one. This normalization prevents the elements of the tensors from becoming out of range for machine precision under successive contractions. 
Alongside the normalized environment, we store $f_i = \sum_j \text{ln}(f_{ij})$. The complete environment can then be reconstructed by multiplying by $\exp(f_i)$ at the end of the process. 

Numerical stability is also critical when evaluating expressions of the form $\frac{\Psi(\textbf{S}')}{\Psi(\textbf{S})}$, which frequently appear in Eq.~\eqref{eq:local_estimator}. In this case, we improve stability by computing $\exp{(\ln{(\Psi(\textbf{S}'))} - \ln{(\Psi(\textbf{S}))})}$ instead, thereby avoiding potential numerical inaccuracies in direct division.

\subsection{On Direct Sampling}

The evaluation of Eq.~\eqref{eq:local_estimator} additionally requires the generation of samples that follow the probability distribution of the PEPS state $p_{\Psi}(\textbf{S}) = \frac{|\Psi(\textbf{S})|^2}{\braket{\Psi|\Psi}}$.
We employ here a scheme that directly generates samples from the PEPS state vector~\cite{Vieijra21_directsampling_PEPS}.
This direct-sampling scheme has several substantial benefits with respect to Markov-chain-based ones~\cite{Liu2017_gradient_fPEPS}.
Firstly, it avoids autocorrelation problems and the necessity for problem-dependent update schemes, as the samples are sampled independently of each other. Secondly, for the same reason, the sample generation can be parallelized. 

It is worth noting that a double-layer tensor network needs to be contracted when employing the direct sampling scheme. 
However, this can be done with a small contract bond dimension $D_c \sim D$, while for expectation values with double-layers a larger contract bond dimension $D_c \sim D^2$ needs to be chosen.
Thus, for small $D$ the double-layer contractions in the direct sampling algorithm are not a dominant contributor to the total computational time due to the other computations necessary for finite PEPS optimizations. 
In the case of large $D$, the double-layer environments can be computed asynchronously.
This will make them slightly outdated, but since these errors are corrected using importance sampling, their accuracy can be easily monitored by making sure that the statistical error for the energy is small enough.
We further comment on this and summarize the direct-sampling scheme in App.~\ref{app:direct sampling of PEPS}.

\subsection{On Variational Upper Bound to the Ground State Energy}
The estimator of Eq.~\eqref{eq:local_estimator} defines a variational upper bound for the ground state energy, provided that a consistent numerical value is assigned to $\Psi(\textbf{S})$.
However, for efficiency in numerical calculations (to reuse boundary-MPS), we may use $L-1$ different approximations for $\Psi(\textbf{S})$, depending on which boundary-MPS we use. 
We denote these different approximations $\Psi(\mathbf{S})_i := E^u[i] \cdot E^l[i+1]$. 
Consequently, the energy expectation value obtained in this manner does not strictly serve as a variational upper bound.

Nevertheless, as long as these different approximations are very close to each other, we can expect to obtain a good approximation of the energy expectation value that is not far from the upper bound.
It is possible to obtain a variational bound on the energy by restricting ourselves to a single approximation $\Psi(\mathbf{S})_i$ throughout the calculation of the expectation value, which is computationally significantly more demanding.
To evaluate the accuracy of the efficient, non-variational scheme, we compare the expectation value obtained in this way to the one obtained to give a variational bound. 
This method was also used in Ref.~\cite{liu2024tensor} to establish variational upper bounds for PEPS.

As an example, the variational energy for the ground state of the $J_1$-$J_2$ model with $J_2 = 0.58$ and $L = 10$ was calculated using a bond dimension $D=6$ PEPS and $\Psi(\mathbf{S})_{i=5}$, with a contraction cutoff of $10^{-4}$ (which translates to $D_c\approx 12$) and $10^5$ samples. The same samples were used for both methods. The fast, inexact method yielded $\braket{H} = -187.856 \pm 0.036$, while the upper bound calculation produced $\braket{H}_\text{bound} = -187.805 \pm 0.027$. 

To estimate the error between these two values, the elementwise difference of the sampled $E^\text{loc}$ was computed as $\braket{E^\text{loc}_\text{bound}(\mathbf{S}) - E^\text{loc}(\mathbf{S})}_{\mathbf{S} \in p_\Psi} = 0.051 \pm 0.013$. This result suggests that during optimization, the ansatz exploited variations in $\Psi(\mathbf{S})_i$ to lower the energy. However, the error is minor, affecting only the fifth significant digit. A more detailed analysis is provided in Appendix \ref{app:true_variational}.


\section{contraction complexity issues in sampling PEPS}
\label{sec:contraction complexity issues in sampling PEPS}

A central mathematical task in the sampling-based schemes for finite PEPS is the contraction of single-layer tensor networks. 
These single-layer networks emerge in the calculation of the wave-function amplitude of a single many-body configuration $\Psi(\textbf{S}) = \braket{\textbf{S}|\Psi}$ as well as in the evaluation of local estimators $\hat{O}_{\text{loc}}(\textbf{S}) = \bra{\textbf{S}}\hat{O}\ket{\Psi}/\Psi(\textbf{S})$. 
However, it was recently discussed in the context of random tensor networks that the contraction of these single-layer tensor networks with the iterative MPS-MPO approach described in Sec.~\ref{sec:finite PEPS and sampling}, can be a task of drastically varying complexity~\cite{chen2024signproblem}.\\

In this section, we are going to analyze the situation first for random PEPS and then focus on the use case for finite PEPS calculations, which involve non-random tensors. 
We first find that the complexity transition found by Schuch \etal~\cite{chen2024signproblem} corresponds to an entanglement transition of the random PEPS.
We then demonstrate that the complexity of contracting the single-layer tensor networks that appear in non-random situations, such as the ground state search of a physical model, is generically less problematic than the ones encountered for random PEPS. 
We point out that, therefore, the naive random PEPS as a starting point of a PEPS optimization should be avoided. 

Note that similar analyses have been done in the context of random quantum circuits \cite{mcginley2024measurement}.

\subsection{Tools for the analysis of contraction complexity}

\subsubsection{Contraction complexity indicators}
\label{sec:contr_compl_issues:compl_indicators}
When contracting a two-dimensional single-layer tensor network, as necessary to obtain $\Psi(\textbf{S})$, with the boundary-MPS method described in Sec.~\ref{sec:boundaryMPS}, we can define several quantities that are helpful in the analysis.\\
Firstly, we can fix a maximal truncation error $\epsilon_{\text{trunc}}$ that we allow within the truncation of the MPS-MPO product. 
We define the boundary-MPS bond dimension $D_c(\epsilon_{\text{trunc}})$, which is necessary to achieve this truncation-accuracy $\epsilon_{\text{trunc}}$ in every step. This quantity can be used to indicate the computational difficulty of the single-layer contraction.\\

We further want to define a quantity that can illustrate the convergence of the scalar value $\psi$ of the contraction of the single-layer tensor network as a function of the cutoff boundary-MPS bond dimension $D_c$. For this, we define
\begin{equation}
    \Delta\psi(D_c, D_c^{\max}) = \frac{\psi(D_c^{\max}) - \psi(D_c)}{\psi(D_c^{\max})},
\end{equation}
where $D_c^{\max}$ is choosen to be substantially larger than $D_c$ ($D_c^{\max} \gg D_c$).
This quantity $\Delta\psi(D_c, D_c^{\max})$, which we refer to as the \textit{relative contraction error} of $\psi$, can be used to establish suitable contract bond dimension $D_c$, to be used in practical calculations. It can additionally be employed to visualize the convergence behavior of different single-layer tensor networks. 

\subsubsection{Geometric Entanglement}
\label{sec:geometric_entanglement}
The geometric entanglement per site, denoted as $S_G$, quantifies the entanglement in a quantum many-body system on a per-site basis. It has previously been used in the MPS context as a tool to visualize phase transitions in~\cite{orus2010visualizing}.
It is defined using the maximum fidelity $\Lambda_{\max}$ between a given pure normalized quantum state $\ket{\psi}$ and the set of all product states $\ket{\phi}$:
\begin{align}
\Lambda_{\max} &= \max_{\ket{\phi}} \left| \bra{\phi} \psi \rangle \right|^2, \nonumber\\
S_G (\ket{\psi}) &= -\frac{1}{N}\log_2 \Lambda_{\max}\;.
\end{align}
This maximum fidelity measures how close the state $\ket{\psi}$ is to being fully separable, and the per-site geometric entanglement $S_G$ provides a straightforward to compute measure of the entanglement in the quantum state which is zero in the case of a product state and one in the case of a maximally entangled state. An efficient scheme to compute the geometric entanglement is described in Sec.~\ref{app:geom_ent}.

\subsubsection{Spectral properties in Vidal gauge}
Generically, tensor networks enjoy so-called \textit{gauge freedom}, from the fact that for every index of the tensor network that is contracted, we can insert a pair of invertible matrices $M\cdot M^{-1} = \mathbb{I}$ without changing the result of the contraction.
In loop-free tensor networks, this gauge freedom plays a crucial role, but even in the context of PEPS a particular gauge choice, which we refer to as \textit{Vidal gauge} is often used in the context of studies using the simple-update technique for optimization\cite{SU_1,SU_2}. This gauge choice can be achieved numerically on PEPS with the use of several algorithms~\cite{SU_1, Belief_prop2023}. In the Vidal gauge the PEPS is represented in a form such that in addition to the local tensors, we have diagonal, non-negative matrices $\Lambda$ on all bonds of the network. The tensors are gauged in such a way as to fulfill a \textit{isometry condition} if the adjacent bond matrices are absorbed into the local tensors
\begin{equation}
\includegraphics[height=5.5cm]{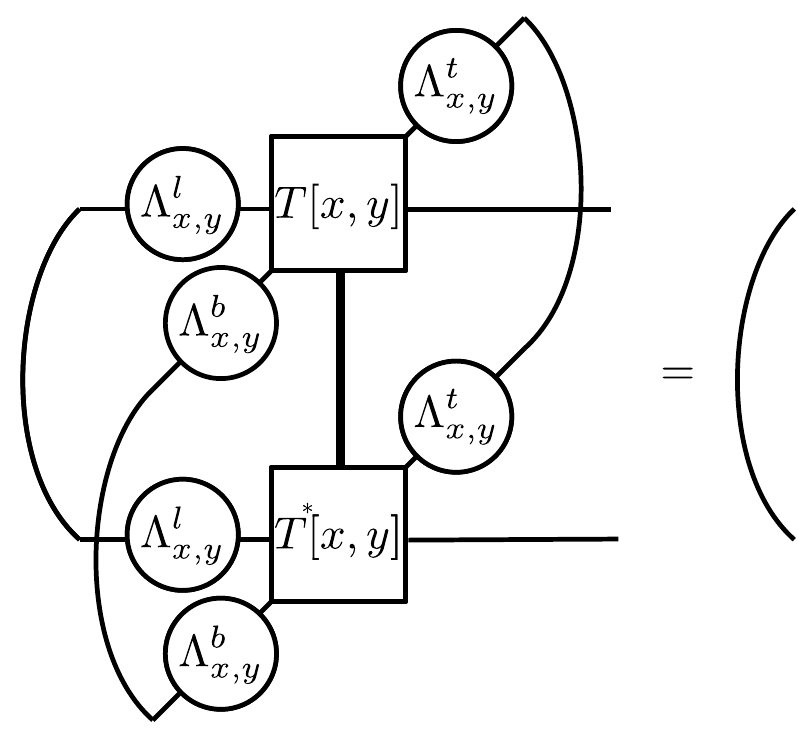}.
\end{equation}
This condition holds in all directions.
Inspired by the fact that for an MPS in the Vidal gauge, the bond matrices carry the Schmidt coefficients~\cite{Vidal2004} we choose to study the properties of the bond matrices of our PEPS in the Vidal gauge. Specifically, we will use the \textit{average spectral entropy} 
\begin{equation}
    \begin{split}
        H_{\text{spec}} &= \frac{1}{N_b} \sum_{i \in \text{bonds}} H_{\text{spec}}^i,\\
        H_{\text{spec}}^i &= \sum_k |S_k|^2 \log(|S_k|^2),
    \end{split}
\end{equation}
where $S_k$ are the entries of the bond matrices of the PEPS in the Vidal gauge. We will see in the following sections that this average spectral entropy can be related to the difficulty of contracting the samples of a PEPS.

\subsection{Random PEPS}
\begin{figure*}[hbt]
    \subfloat[]{\label{fig:random_peps:shift:Sg}\includegraphics[width=0.45\textwidth]{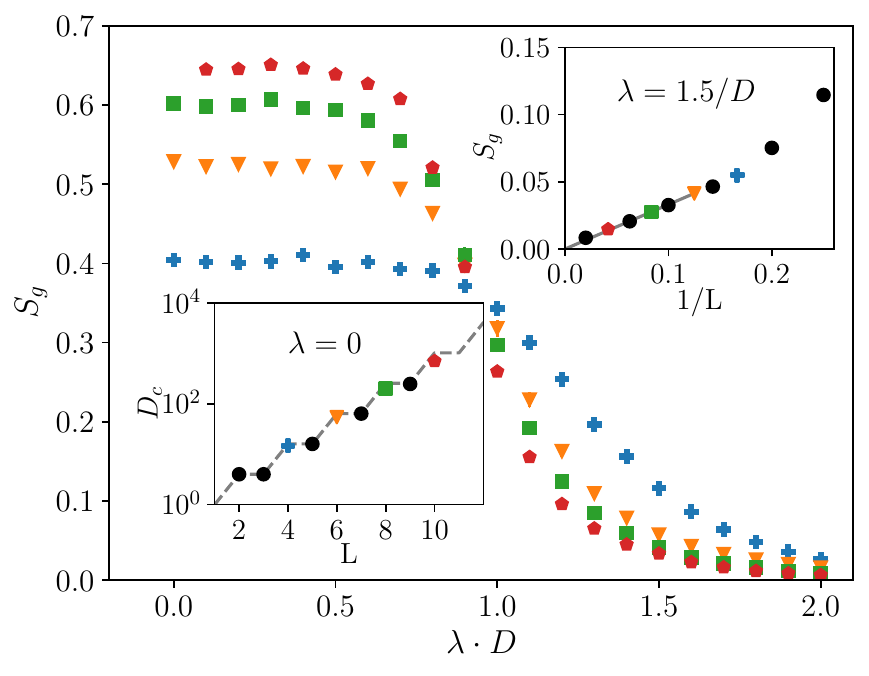}}
    \subfloat[]{\label{fig:random_peps:spectrum:Sg}\includegraphics[width=0.45\textwidth]{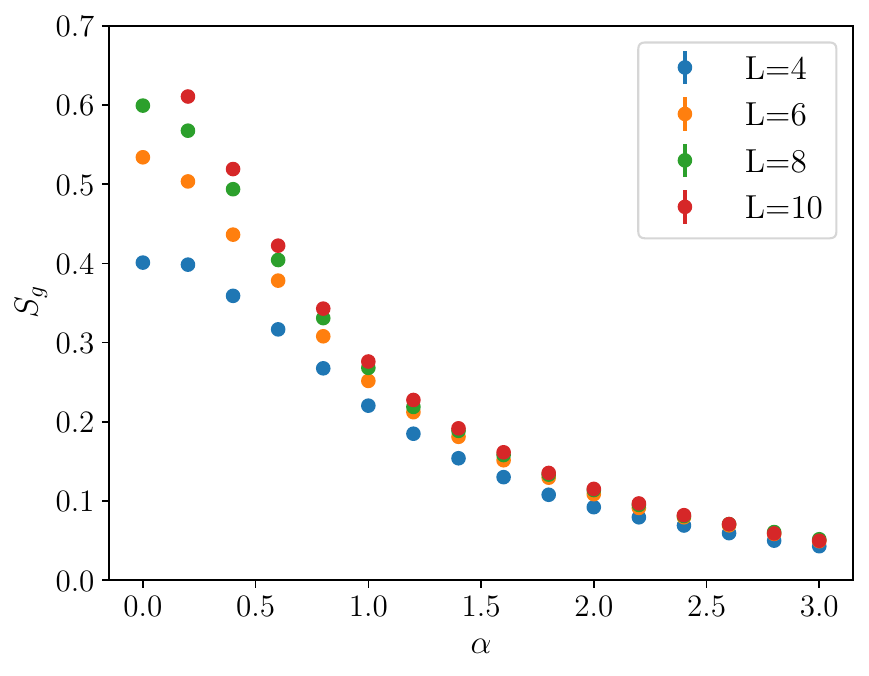}}
    
    \caption{Geometrical entanglement $S_g$ for random finite PEPS with bond dimension $D=4$ and different system sizes for (a) a normally distributed tensor shifted by a constant $\lambda$ and (b) a normally distributed tensor with algebraically decaying diagonal matrices $S_{i,j}=\delta_{i,j} 1/i^\alpha$ multiplied on each nonphysical link. The bottom left inset in (a) shows the dynamically determined contracted bond dimension $D_c$ of the boundary MPS with a cutoff of $10^{-4}$ at different system sizes for $\lambda=0$, where the dashed line represents the maximal possible bond dimension for an MPS of size $L$. The top right inset in (a) presents $S_g$ versus system size, demonstrating that the geometrical entanglement approaches zero in the infinite system size limit, indicating that the PEPS collapses into a product state.}
    \label{fig:random_peps:spectrum}
\end{figure*}%

We begin our analysis from random PEPS since their wave function amplitudes $\Psi(\textbf{S})$ for any sample $\textbf{S}$ correspond to the random two-dimensional tensor networks investigated in~\cite{chen2024signproblem}.
A central result of that study is that the complexity of the contraction of a Haar-random two-dimensional tensor network with the boundary-MPS method can be changed when shifting the mean value of the random tensors $\lambda$ towards positive entries. 
At a mean value of $\lambda = 1/D$, where $D$ denotes the bond dimension of the local tensors in the two-dimensional network, the complexity of the approximate contraction changes drastically. 
The change in complexity is measured e.g. with the bipartite entropy of the boundary MPS used for the contraction or equivalently the boundary MPS bond dimension necessary.

We are now investigating the properties of random PEPS to see if we can find a physical quantity that serves as a predictor for the complexity of contracting its samples.
As a first step, we are going to again shift the mean of the entries of the random PEPS by $\lambda$ and examine its entanglement via the geometric entanglement, cf. Sec.~\ref{sec:geometric_entanglement}. The results are shown in Fig.~\ref{fig:random_peps:spectrum}\subref{fig:random_peps:shift:Sg}. 
We notice that at large shifts $\lambda$, the geometric entanglement becomes very small and the random PEPS represent essentially a product state. 
By extrapolation in system size, we show that in infinite systems this seems to be the case.
If we now decrease the shift $\lambda$ we notice an increase in the geometric entanglement. At the value of $\lambda = 1/D$, we find a transition in the geometric entanglement, when scaling the size of the system under consideration. At shift values below this transition, we find that the geometric entanglement plateaus to a value set by the system size. The value of the plateau converges to 1 (the maximum geometrical entanglement) as the system size grows.

This suggests that random PEPS go through an entanglement phase transition (as measured here by the geometric entanglement) as a function of the value of the mean-value shift $\lambda$ which coincides with a complexity transition for the task of contracting the corresponding samples as established in~\cite{chen2024signproblem}.\\

This drastic change in the computational difficulty of contracting the random-single-layer networks suggests an important question.
Is such a transition in the difficulty of the contraction a generic feature, when changing the tensors of the network?
This is an important, practical consideration for sampling-based PEPS calculations because we will continuously manipulate the PEPS tensors during the ground state search.
In this ground state search procedure, should one expect to generically hit a complexity phase transition, at which point the calculation of observables becomes very hard and the method stops working in practice?\\
We investigate this question in several steps. First, we are again considering random PEPS, which we now manipulate in a different way. Instead of shifting the mean value of the entries, as done previously, we multiply the virtual legs of every local tensor in the PEPS with diagonal matrices $S(\alpha)_{i,j}=\delta_{i,j} 1/i^\alpha$, whose entries decay as a function of $\alpha$
\begin{equation}
    T(\alpha) = (S(\alpha)^{\otimes 4})T \; ; \, T(\alpha) = \begin{gathered}
        \includegraphics[height=2.5cm]{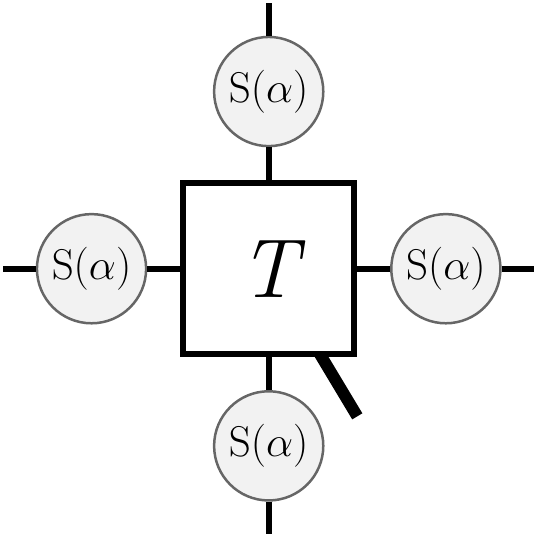}
    \end{gathered}.
\label{eq:A_alpha}
\end{equation}
We show the geometric entanglement of the resulting PEPS as a function of the decay-parameter $\alpha$ in Fig.~\ref{fig:random_peps:spectrum}\subref{fig:random_peps:spectrum:Sg}. We observe a smooth increase of the geometric entanglement as the decay-parameter $\alpha$ decreases. 
This suggests a smooth entanglement-crossover as a function of $\alpha$ and stands in stark contrast to the entanglement transition shown in Fig.~\ref{fig:random_peps:spectrum}\subref{fig:random_peps:shift:Sg}. 
We take this as first, tentative evidence that the drastic complexity transition observed for random tensors with a shifted mean $\lambda$ is not generic.\\

Furthermore, we point out, that the contraction complexity of samples of a PEPS state vector cannot be entirely due to its entanglement properties. 
To illustrate this, we consider a PEPS $\ket{\phi}$, made up of a single-layer network to which we attach physical legs, without connecting them to the single layer. 
\begin{equation}
    \ket{\phi} = \begin{gathered}
    \includegraphics[height=3.5cm]{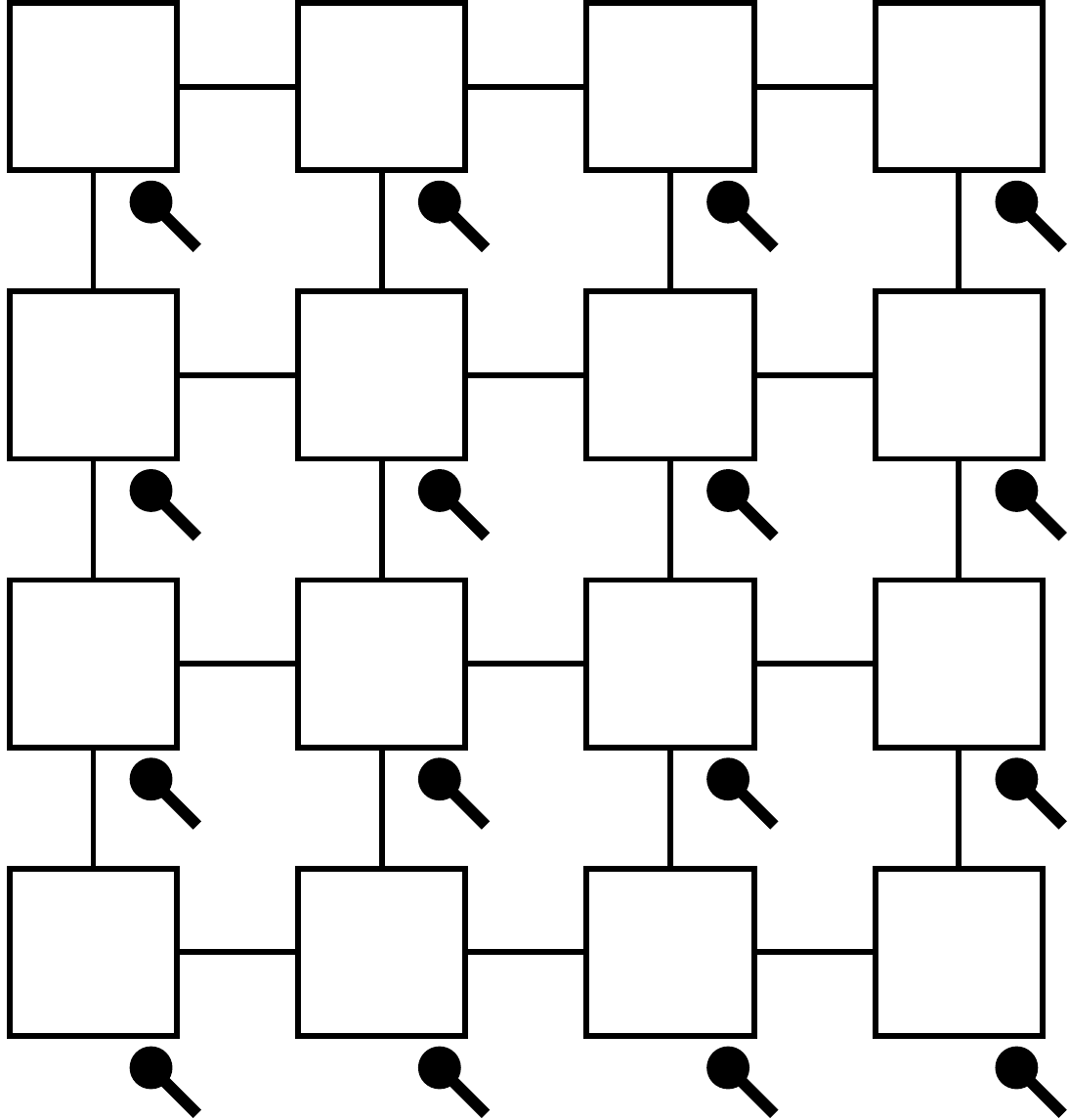}
    \end{gathered}
\end{equation}

In this construction, every basis vector of the resulting many-body state vector has the same coefficient.
Such a situation represents a product state, such that there is no entanglement for any bipartition. However, if we choose a single layer in the above construction that is hard to contract, every sample of this PEPS is hard to contract even though the PEPS represents a product state.

\subsection{Towards physical PEPS}

\begin{figure*}[t]
    \subfloat[]{\label{fig:J1J2:DeltaPsi_asfuncof_alpha}\includegraphics[height=0.34\textwidth]{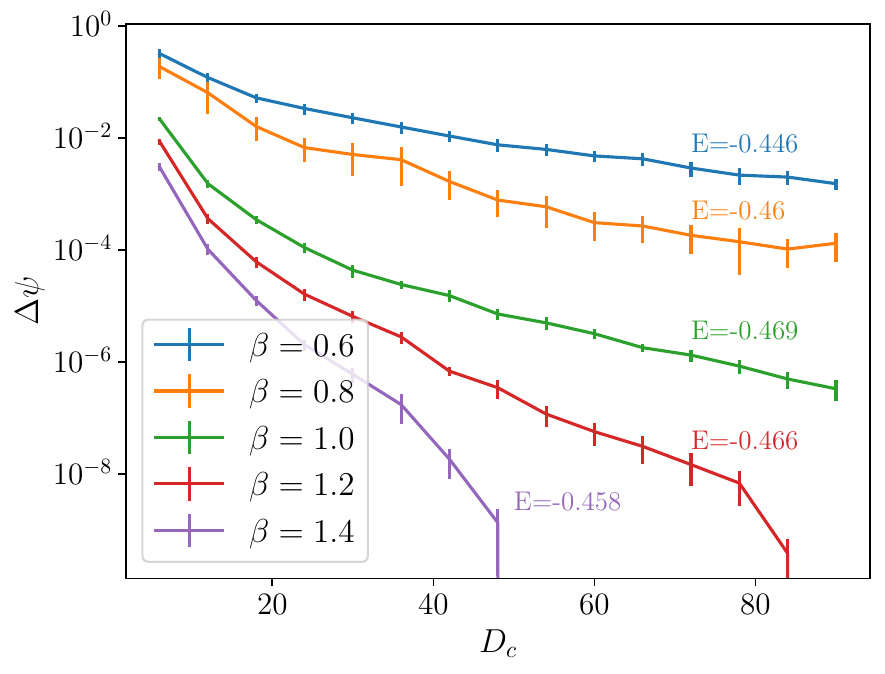}}
    \subfloat[]{\label{fig:J1J2:candidate_quant_asfuncof_alpha}\includegraphics[height=0.34\textwidth]{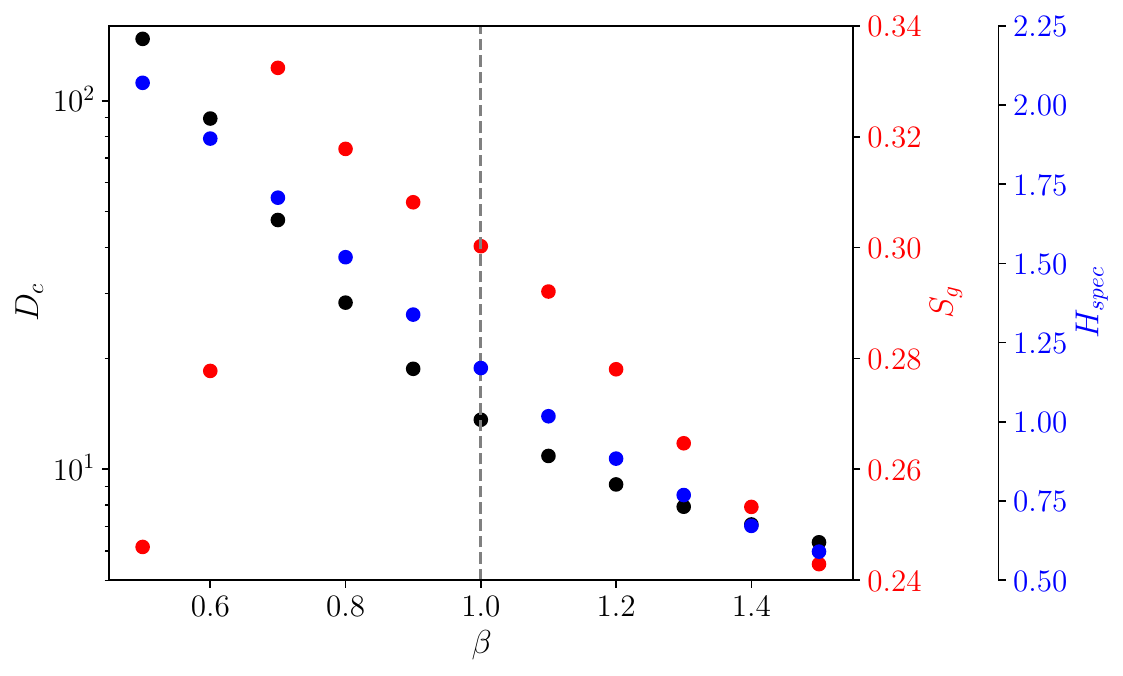}} 
    \caption{
    (a) The relative contraction error, $\Delta\psi(D_c, D_c^{\max} = 200)$ is plotted against the contraction dimension $D_c$ for various values of $\beta$, with which the Vidal spectra were modified modified as $\{ S \}_{\text{bonds}} \mapsto \{ S^{\beta} \}_{\text{bonds}}$. The energy of the resulting states is also recorded. (b) The contraction dimension $D_c$, geometrical entanglement $S_g$, and average spectral entropy $H_\text{spec}$ are plotted as functions of $\beta$. The results show that $H_\text{spec}$ is correlated with $D_c$, whereas $S_g$ does not exhibit a direct correlation with $D_c$.}
    \label{fig:J1J2_fig1}
\end{figure*}%

\begin{figure*}[t]
    \subfloat[]{\label{fig:J1J2:energy}\includegraphics[height=0.23\textwidth]{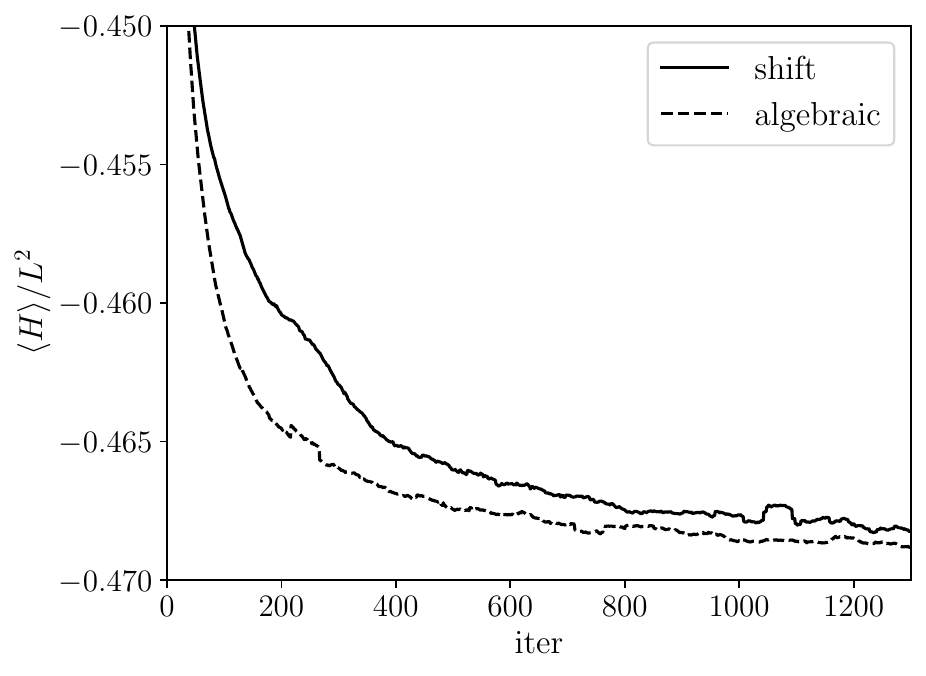}}
    \subfloat[]{\label{fig:J1J2:CD_SG_HS}\includegraphics[height=0.23\textwidth]{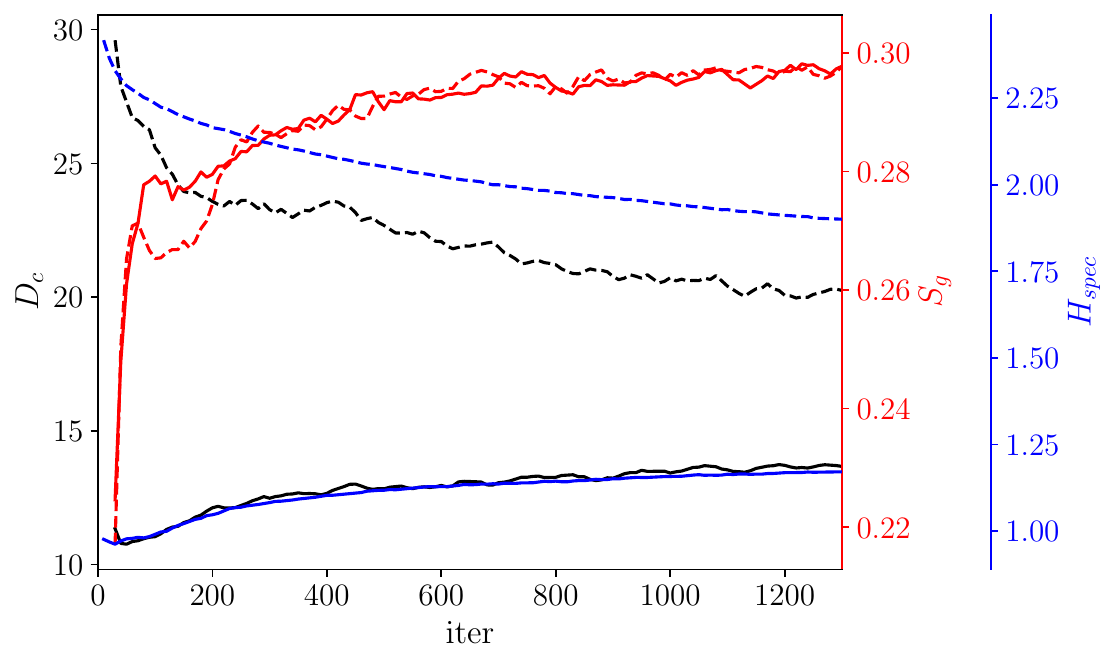}}
    \subfloat[]{\label{fig:random_peps:spectrum:Dc_Sg}\includegraphics[height=0.23\textwidth]{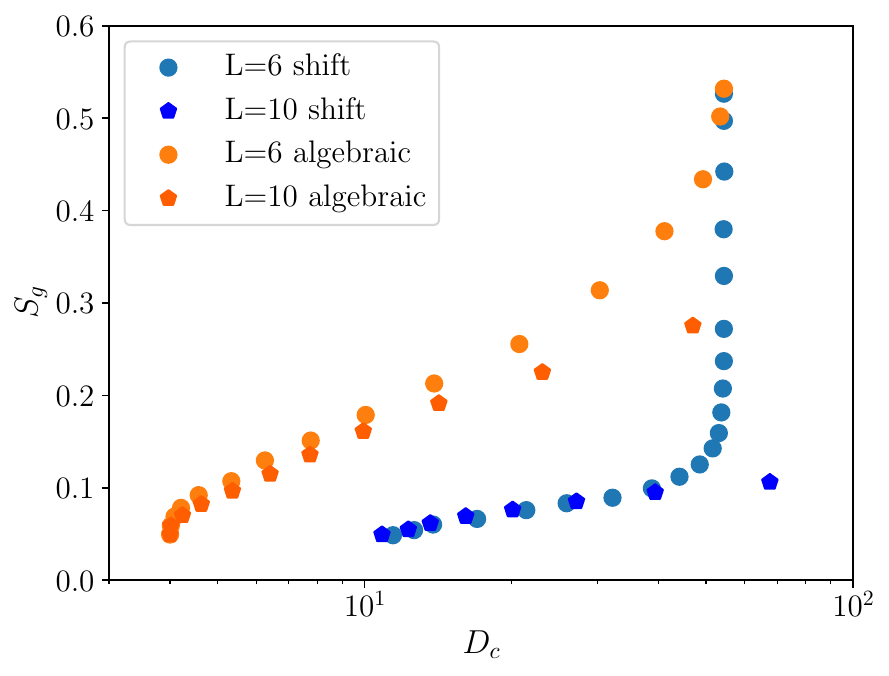}}
    \caption{Different quantities calculated during the optimization of a PEPS initialized (solid line) with a shift $\lambda=2$ and (dashed line) an algebraically decaying spectrum $\alpha=1.5$. (a) Energy per site $\langle H \rangle / L^2$ versus optimization iterations. (b) Evolution of the contraction bond dimension $D_c$ (black), geometrical entanglement $S_g$ (red), and spectral entanglement $H_s$ (blue) during the optimization. (c) Contract environment Dimension $D_c$ for randomly sampled finite PEPS that was either shifted or multiplied times an algebraic decaying spectra plotted against its geometrical entanglement.} 
    \label{fig:J1J2}
\end{figure*}%
We now move to investigate PEPS that are obtained during the ground state search of generic physical models.
We aim to find a way to characterize and predict the contraction complexity of the samples of the PEPS obtained in these scenarios.\\

To start we consider the ground state PEPS approximation obtained for the $J_1-J_2$ model, cf. benchmarks in Sec.~\ref{sec:applications}.
We will now describe a way to create from this ground-state PEPS a family of low-energy PEPS, which is well suited for our study of the contraction complexity.
We choose to represent the ground state PEPS in the Vidal gauge, cf. Sec.~\ref{sec:contr_compl_issues:compl_indicators}, which allows us to generate from it a set of new PEPS by the following procedure. 
We can take the set of spectra on the bonds of the ground state PEPS in the Vidal gauge and take them to the power of $\beta$,
\begin{equation}
    \{ S \}_{\text{bonds}} \mapsto \{ S^{\beta} \}_{\text{bonds}},
\end{equation}
such that for every choice of $\beta$ we obtain a unique new PEPS. 
For the family of PEPS that we have obtained in this way we now investigate the contraction complexity with the use of the relative contraction error $\Delta\psi(D_c, D_c^{\max})$. 
In Fig.~\ref{fig:J1J2:DeltaPsi_asfuncof_alpha} we show for several choices of $\beta$, that such a manipulation yields PEPS with different contraction complexities for their samples.
In fact it is quite clear from Fig.~\ref{fig:J1J2:DeltaPsi_asfuncof_alpha}, that a manipulation that flattens the spectra ($\beta < 1$) yields a PEPS whose samples are more difficult to contract. 
Conversely, a steepening of the spectrum ($\beta > 1$) yields PEPS whose samples are cheaper to contract. 
This suggests that the spectrum in the Vidal gauge might be related to the contraction complexity of physical PEPS.
Remarkably, all of the PEPS that we generate in this way are energetically close to the ground state, even if the contractability of their samples varies drastically.

For this reason, we choose this family of low-energy PEPS, parameterized by the single parameter $\beta$, and try to find a quantity that can predict how difficult it is to contract the samples of a given PEPS. 
In Fig.~\ref{fig:J1J2:candidate_quant_asfuncof_alpha} we show several candidate quantities as a function of $\beta$ as well as the cutoff boundary-MPS bond dimension $D_c$ which is necessary to maintain an error of $\epsilon_\text{trunc} = 10^{-4}$ during the contraction procedure. 
We first point out that the geometric entanglement does not seem to be a good predictor of the cost of the single-layer contractions, as might have been conjectured based on the investigations of random PEPS in the previous section. 
In fact for small $\beta$ the cutoff bond dimension $D_c$ increases, while the geometric entanglement decreases. 
This should further be taken as evidence that random PEPS do not generically share properties with physical PEPS. 
However, motivated by the fact that a manipulation of the Vidal-spectra of the PEPS seems to have a large impact on the contractability of their samples, cf. Fig.~\ref{fig:J1J2:DeltaPsi_asfuncof_alpha}, we also consider the average spectral entropy $H_\text{spec}$ of the PEPS at different $\beta$ in the Vidal gauge. Fig.~\ref{fig:J1J2:candidate_quant_asfuncof_alpha} shows that this quantity seems to be correlated with the difficulty of contraction.


In order to further substantiate the above finding that the spectral properties in the Vidal gauge are strongly correlated with the cost of contracting samples of a physical PEPS, we look at another family of PEPS, namely those obtained during the ground state search with imaginary time evolution, cf. Sec.~\ref{sec:imaginary_schrodinger_eq}. 
To this end, we have performed two optimization runs with different initial tensors. 
We have chosen a set of mean-shifted random tensors ($\lambda = 2$) as a starting PEPS, as well as a set of random PEPS that were manipulated according to Eq.~\eqref{eq:A_alpha} with $\alpha=1.5$. 
Fig.~\ref{fig:J1J2} shows the results of these two ground-state searches. Fig.~\ref{fig:J1J2:energy}  shows the energy of the PEPS during the optimization.
We find that both optimizations converge to a similar energy. However, the random tensors manipulated according to Eq.~\eqref{eq:A_alpha} serve as a more effective initial choice for the PEPS. This approach leads to convergence in fewer iteration steps. Additionally, each iteration step is faster to compute due to the lower contraction dimension $D_c$.

This is a direct consequence of the relation between geometrical entanglement $S_g$ and contract bond dimension $D_c$ for the shifted tensors. In Fig.~\ref{fig:random_peps:spectrum:Dc_Sg} these two quantities are plotted against each other for different values of shift $\lambda$ and algebraic decay $\alpha$. One can see that to obtain a state distinct from the product state for the shifted initialization, a large contract bond dimension is needed, while for the algebraic decaying random PEPS, this is not the case.

More importantly, we find again that the average spectral entropy $H_\text{spec}$ is strongly correlated with the contraction cost of the samples of the PEPS during the optimization. This is true for both choices of initial tensors.
We close this section by concluding that, firstly, the complexity/entanglement phase transition as they can occur in random tensors-network/PEPS does not seem to be a generic feature of low energy PEPS of physical models, which can be considered as a validation of the applicability of the sampling PEPS methods. Secondly, we find that the spectral properties of the PEPS in the Vidal gauge are correlated with the contraction difficulty of its samples.

\section{Applications}
\label{sec:applications}
Now that we have set contraction complexity issues on firm(er) grounds, we proceed to use the finite PEPS framework discussed in Sec.~\ref{sec:finite PEPS and sampling} to investigate several challenging situations. We start by considering the well-studied $J1-J2$ model~\cite{J1J2classic,J1J2classic2,LIU2022_fPEPS} and use this as a starting point to compare the finite PEPS results to those of tree-tensor-network~\cite{Shi06_TTN,Tagliacozzo09_TTN, Rizzi17_FQH_trees,Rizzi19_anthology} in the light of the different entanglement scaling of the two ansätze. 
We then move to an investigation of a Hamiltonian hosting a chiral spin liquid as its ground state; a notoriously difficult quantum state for PEPS~\cite{Wahl13_chiralPEPS, Dublain15_nogo}. This has so far not been investigated with finite PEPS, but recently experienced advances with infinite PEPS optimized variationally with the help of automatic differentiation~\cite{Hasik22_CSL}. Finally, we show how powerful the described method can be for systems with long-range interactions by applying it to describe ground states of Rydberg atom arrays~\cite{labuhn2016_2dRysbergIsing, ebadi2021_rydbergExperiment}, a very prominent platform for quantum simulation.

\subsection{Entanglement scaling and comparison to Tree Tensor Networks: $J_1-J_2$-model}
\label{sec:J1J2}

This section aims to demonstrate that finite PEPS optimized from scratch using sampling TDVP can represent the ground states of the $J_1$-$J_2$ model on the square lattice
\begin{equation}
    \hat{H} = J_1\sum_{\langle i,j \rangle} \vec{S}_i \vec{S}_j + J_2\sum_{\langle\langle i,j \rangle\rangle} \vec{S}_i \vec{S}_j.
\end{equation} 
Previous studies have extensively analyzed the ability of finite PEPS to capture the different phases of this model~\cite{Liu2018_fPEPS_SL, Liu2021_accurate_fPEPS, LIU2022_fPEPS}. 
Here, we verify certain points of the established phase diagram and benchmark a specific point against TTNs (tree tensor networks) computed using \cite{TTN.jl}. 

\begin{figure}[t]
    \subfloat[]{\label{fig:J1J2_optims:phases}\includegraphics[width=0.45\textwidth]{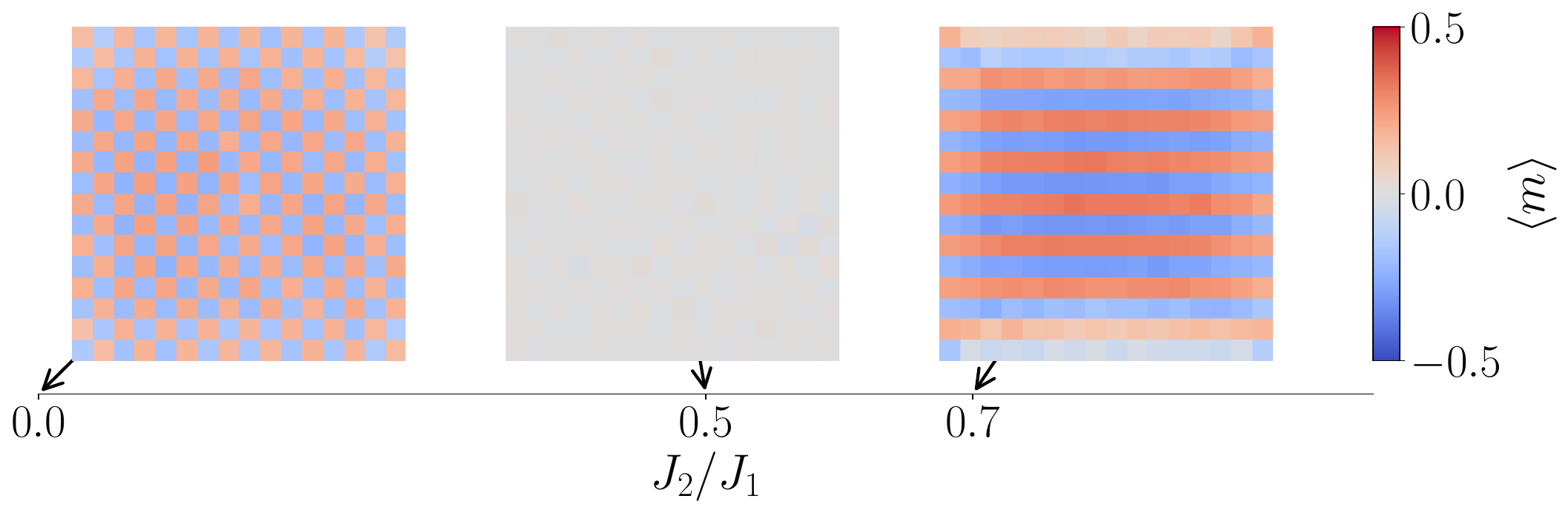}}\\
    \subfloat[]{\label{fig:J1J2_optims:trees_vs_fpeps}\includegraphics[width=0.45\textwidth]{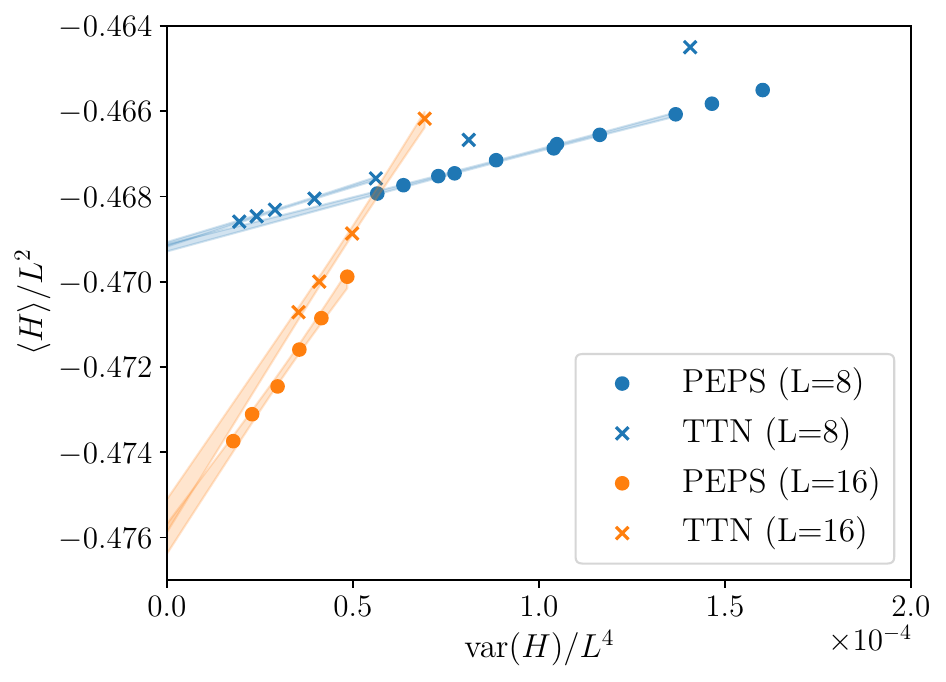}}
    \caption{(a) Average magnetization per site for three distinct phases of the $J_1$-$J_2$ phase diagram. (b) Variational energies plotted against energy variances at $J_2/J_1 = 0.58$ for system sizes $L=8$ and $L=16$. The data points shown as circles correspond to PEPS states with bond dimension $D=8$, optimized using a cutoff of $10^{-4}$, and subsequently evaluated with a more stringent cutoff of $10^{-6}$. Each point represents a PEPS state at specific iterations during optimization: iterations \{350, 410, 490, 650, 790, 870, 1250, 1710, 1950, 3000, 4000\} for $L=8$, and \{830, 1100, 1500, 2420, 4000, 9020\} for $L=16$. A detailed analysis determining the appropriate cutoff value is provided in App.~\ref{app:true_variational}. Data points represented by crosses indicate Tree Tensor Network (TTN) results with bond dimensions $D = \{50, 100, 150, 200, 250, 300, 350\}$ for $L=8$ and $D = \{50, 100, 150, 200\}$ for $L=16$. Dashed lines indicate linear fits employed for extrapolating energies to the zero-variance limit.}
    \label{fig:J1J2_optims}
\end{figure}

We find the expected patterns of the local magnetization, as shown in Fig.~\ref{fig:J1J2_optims:phases}. In the Néel antiferromagnetic phase, we find a checkerboard pattern, while at large values of $J_2/J_1$ we find a stripe-ordered pattern. In the intermediate region, we find vanishing local magnetization. 
Let us note that the magnitude of the local magnetization is smaller at the edges of the system due to the impact of a smaller number of neighbors and, hence, larger quantum fluctuations dressing the order.

To assess how the accuracy of finite PEPS scales with system size and bond dimension, we refer to Fig.~\ref{fig:J1J2_optims:trees_vs_fpeps}. In this figure, the variance per site is plotted against the expectation value of the energy for both tree tensor networks and finite PEPS as is often done in the Neural Quantum state literature \cite{nomura2021helping}. Linear extrapolation to the zero-variance limit shows the estimated energy value for trees and PEPS match. We note that to reach the regime of truly linear relations between energy and variance, larger bond dimensions might need to be considered, especially for the tree tensor networks on larger systems. 

PEPS fulfill the boundary law of entanglement entropy and are thus a good ansatz for any system size. In contrast, TTN, which does not fulfill the boundary law, requires a drastically increasing bond dimension to capture the same amount of entanglement as a PEPS at larger system sizes. However, they can still be an effective ansatz in smaller systems.

As expected, at a small system size of $L = 8$, TTNs with bond dimension 300 achieve 45\% lower energies than PEPS with bond dimension 8, taking the extrapolated energy $E_\text{min}$ as a reference value $\frac{\braket{H}_\text{TTN300} - E_\text{min}}{\braket{H}_\text{PEPS8} - E_\text{min}} = 0.45$. However, at a larger system size of $L = 16$, this trend reverses due to the more favorable entanglement scaling of the PEPS ansatz. In this case, the PEPS ansatz attains 39\% lower energy, with $\frac{\braket{H}_\text{PEPS8} - E_\text{min}}{\braket{H}_\text{TTN200} - E_\text{min}} = 0.39$.

It is important to note that the bond dimension for the tree tensor networks with $L = 16$ was limited to $D = 200$ due to the memory constraints of the V100 GPU with 32 GB of memory.

\subsection{Chiral spin liquid}
\begin{figure*}[t]
    \subfloat[]{\label{fig:CSL_energy}\includegraphics[height=0.32\textwidth]{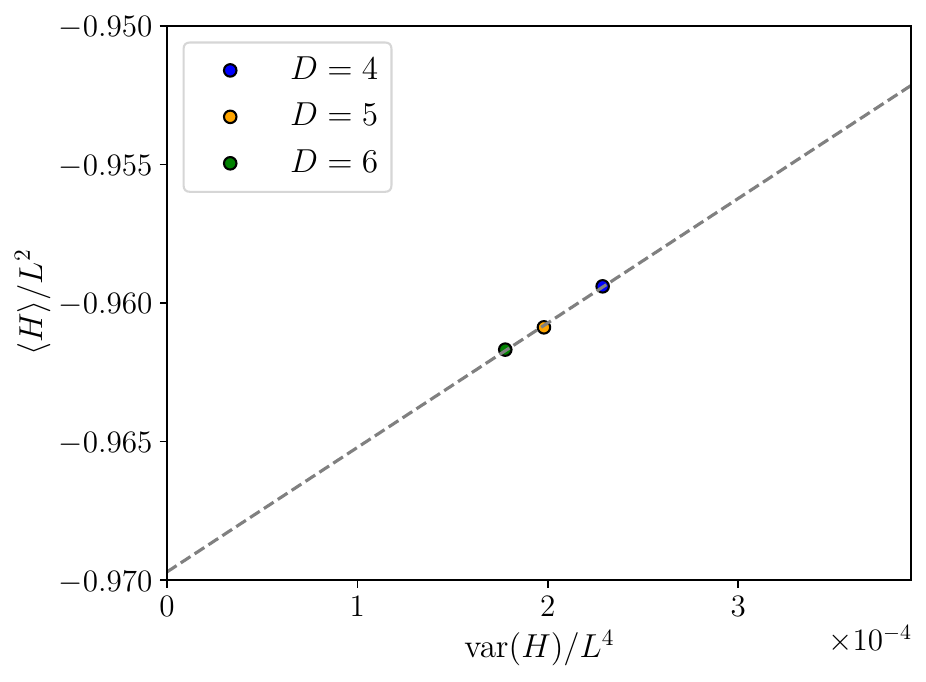}}
    \subfloat[]{\label{fig:CSL_edge}\includegraphics[height=0.32\textwidth]{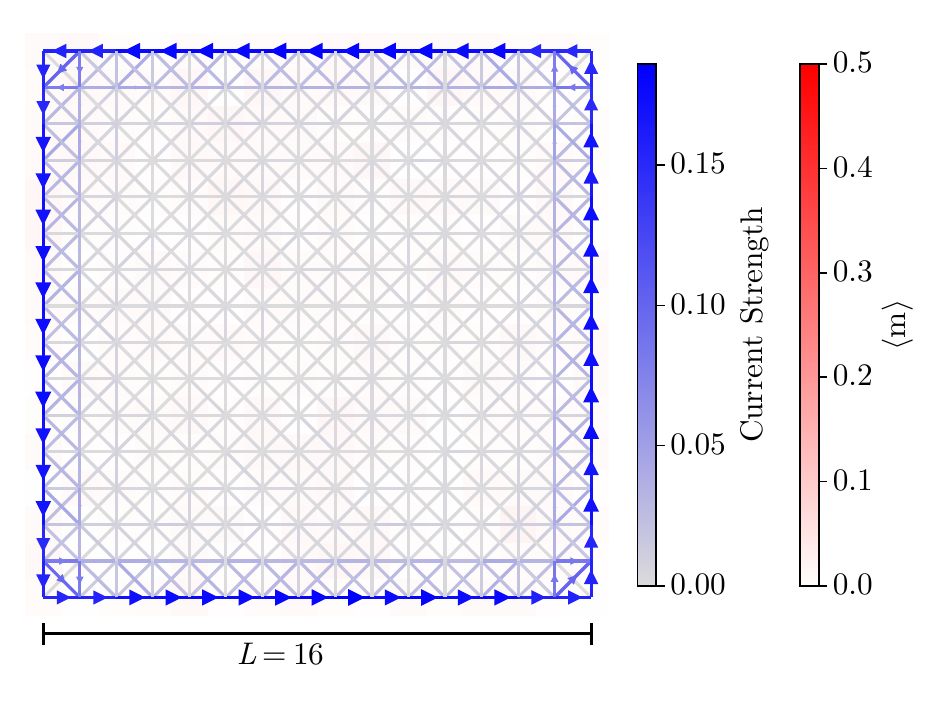}} 
    \caption{Results for the chiral spin liquid. (a) Energy extrapolation using the variance of the results for different bond dimensions. (b) Illustration of observables on a $16\times 16$ lattice. On the vertices, we show the local magnetization, which is homogeneously very small ($m^2 \approx 7\cdot10^{-5}$). On the edges, we show the current. We find a current on the edge of the lattice with preferred chirality.}
    \label{fig:CSL}
\end{figure*}%
Several recent experiments have made strides toward the realization of chiral topological many-body quantum states on quantum simulation platforms~\cite{clark2020FQH, Greiner2023FQH, Jochim24_FQH}. 
These pioneering works involve very few atoms and substantial efforts are being employed towards realizations with increasing numbers of constituents.
For this effort, a faithful simulation of the experimental situation on classical computers - the generation of so-called \textit{digital twins}- is crucial for benchmarking purposes.
The finite PEPS are the natural candidate for this task, as they share the expected entanglement scaling of the chiral topological states~\cite{Haque07_EE_FQH}, that are targeted in the experiments.

As a proof of principle, we examine the Hamiltonian proposed by Nielsen \etal~\cite{nielsen2013local_FQH}, 
\begin{equation}
    \hat{H} = J_1\sum_{\langle i,j \rangle} \vec{S}_i \vec{S}_j + J_2\sum_{\langle\langle i,j \rangle\rangle} \vec{S}_i \vec{S}_j + i\lambda \sum_{\Box} (P_{ijkl} - P^{-1}_{ijkl}),
    \label{eq:CSL_Model}
\end{equation}
which has a chiral spin liquid ground state at $J_1 =2\cos(0.06\pi)\cos(0.14\pi)$, $J_2 = 2\cos(0.06\pi)\sin(0.14\pi)$ and $\lambda = 2\sin(0.06\pi)$. The operator $P_{ijkl}$ is defined for all plaquettes of the system and acts as cyclic permutations on the local Hilbert spaces. The corresponding terms in the Hamiltonian explicitly break time-reversal symmetry. 

Recent works using an iPEPS ansatz have shown that in the thermodynamic limit, chiral topological ground states of local Hamiltonians~\cite{Hasik22_CSL, Weerda24_FQH} can be successfully approximated using the iPEPS ansatz if variational optimization is employed~\cite{Corboz16_variational,Vanderstraeten16_gradient,Liao19_AD_PEPS,Weerda24_variPEPS}. 
For finite system sizes, as relevant for the next generation of mesoscopic cold-atom experiments, however, such a numerical demonstration is to this date lacking.
We show that with the approach for the finite PEPS detailed in this paper, we can indeed find chiral topological states as ground states.

To start, we show in Fig.~\ref{fig:CSL_energy} the extrapolation of the energy density against the variance, cf. Sec.~\ref{sec:J1J2}, for a $16 \times 16$ square lattice with open boundary conditions. 
The estimated energy expectation value of $\langle H \rangle = -0.9697 \pm 0.0005$ is quite close to the result obtained for infinite systems using translationally invariant infinite PEPS where it is found numerically to be close to~$-1$~\cite{Hasik22_CSL}. One can get even closer to the expectation value found in the thermodynamic limit by extracting the energy expectation value for the bulk of the finite PEPS. For the terms of the Hamiltonian corresponding to the center $8 \times 8$ lattice sites of our $16 \times 16$ lattice, we obtain an energy of $E_{\text{bulk}} = -0.99335 \pm 0.00439$.
Next, we focus on local observables that could act as local order parameters. 
We plot $m_i = |\langle \vec{S}_i \rangle |$ for all sites of the lattice in Fig.~\ref{fig:CSL_edge}. 
We find that the local magnetization becomes very small $m^2 \approx 7\cdot 10^{-5}$, suggesting vanishing magnetization in the state.

To briefly investigate the chiral nature of the state, we focus on its edge. For chiral topological states on finite systems with open boundary conditions, we expect chiral behavior at the edges, as demonstrated e.g. in~\cite{Rizzi17_FQH_trees, banerjee2023electromagnetic}. 
In~\cite{Khomskii08_edgecurrent}, it was shown that for Mott insulators, loop currents around triangles can be shown to have the form
\begin{equation}
    \mathcal{I}_{ij,k} \sim \frac{\textbf{r}_{ij}}{|\textbf{r}_{ij}|}\vec{S}_i \cdot ( \vec{S}_j \times \vec{S}_k),
\end{equation}
where $\mathcal{I}_{ij,k}$ denotes the current contribution of the loop current around the triangle consisting of sites $i,j$ and $k$ along the edge connecting site $i$ and $j$.
This quantity is promising for our investigation as the triple-product is invariant under $SU(2)$ spin rotations, which is required for its expectation value not to vanish for the chiral spin liquid. 
The Hamiltonian in Eq.~\eqref{eq:CSL_Model} contains next-nearest neighbor spin interactions (which can be interpreted as perturbatively arising from next-nearest neighbor hopping terms in an underlying fermionic model). 
Thus, in order to investigate the contribution of the loop current in our quantum state, we sum over all triangles of our system, consisting of two pairs of nearest neighbors and one pair of next-nearest neighbors. 
The results are shown in Fig.~\ref{fig:CSL_edge}.
We find that the circulating currents in the bulk cancel out, such that no net current is found, while at the edges, such a cancellation does not happen, such that a chiral current around the edge emerges, as expected.  

\subsection{Rydberg atom arrays and long-range interactions}

Long-range interactions are present in many different scenarios of interest in modern condensed matter- or atomic physics. Prominent examples include Coulomb-interactions ($U_C \sim \frac{1}{r}$) in ab initio electron systems, dipolar-interactions in cold quantum gases ($U_d \sim \frac{1}{r^3}$)~\cite{Lahaye_2009_dipolar_gases} as well as van-der-Waals interaction in Rydberg atom arrays ($U_{vdW} \sim \frac{1}{r^6}$)~\cite{labuhn2016_2dRysbergIsing, ebadi2021_rydbergExperiment}.

Traditionally, the treatment of such long-range interactions in the context of tensor-network methods (specifically in two dimensions) has been challenging in non-sampling-based approaches.
This is because, for a system of $n$ sites, the evaluation of every single term necessitates an evaluation of the order of $n^2$ terms, each one of which might be hard to evaluate depending on the specific method used. Therefore, in almost all cases, interactions have been truncated to a few neighbors.
One can make substantial progress by employing projected entangled-pair operator methods~\cite{ORourke2018_PEPO1,ORourke2020_PEPO2}, which come at the cost of approximating the long-range interaction.

In stark contrast to these challenges, within the sampling-based approach used in this paper, the evaluation of long-range interactions becomes trivial, as long as the interaction is diagonal in the computational basis. As already briefly noted in Sec.~\ref{sec:boundaryMPS}, the evaluation of the expectation value of any operator $\hat{O}_{diag}$ diagonal in the computational basis, requires no additional numerical calculations beyond the wave function amplitude for every sample
\begin{equation}
    \bra{\textbf{S}}\hat{O}_{diag}\ket{\Psi} = O_{\textbf{S}\textbf{S}}\Psi(\textbf{S}).
\end{equation}
This is true irrespective of how local the operator in question acts, which is why we can treat these long-range interactions cheaply, which we exploit in the following.
Additionally, the direct sampling procedure does not rely on a specific update scheme for generating the relevant samples of a state vector. The approach used in this paper is ideally suited for the investigation of models with long-range interactions in the finite PEPS framework.

\label{sec:Rydbergs}
\begin{figure}[t]
    
    \includegraphics[width=0.49\textwidth]{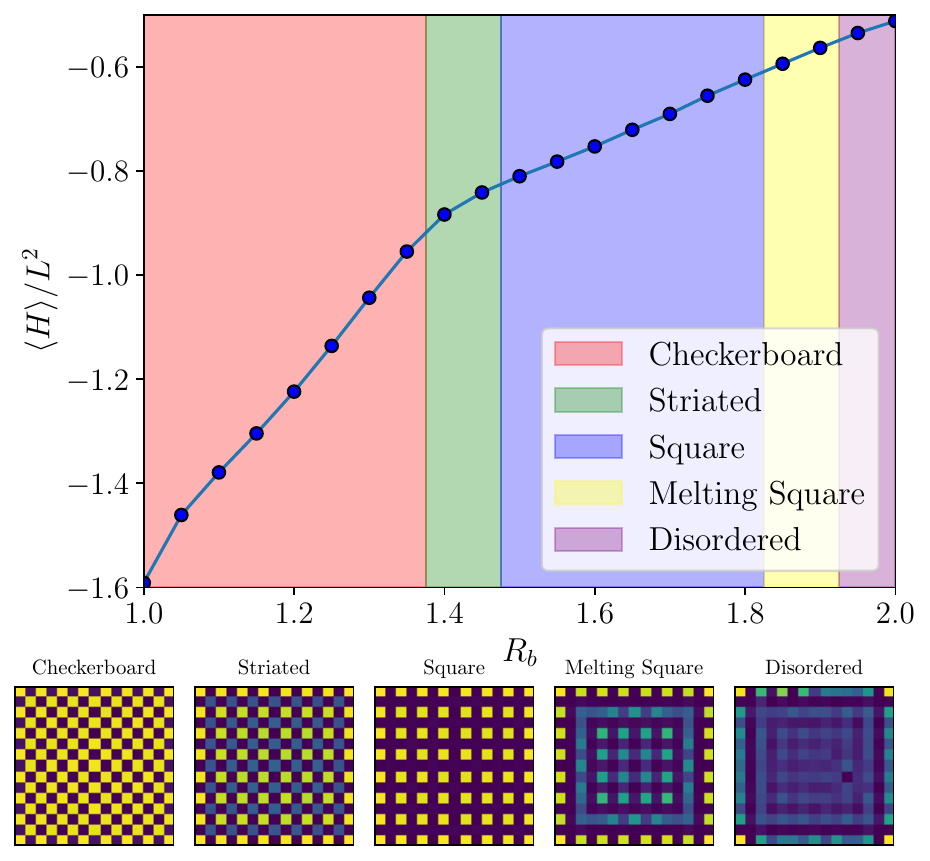}
    \caption{Results on the Rydberg atom arrays. We show the energy of the configurations we found for the fixed detuning $\Delta = 3$ and various values of $R_b$. We highlight the local Rydberg-density patterns for the different phases below.}
    \label{fig:Rydbergs}
\end{figure}
To demonstrate this, and keeping with the theme of digital twins of current cold atom experiments, we are investigating the phases of a Rydberg atom array on a square lattice. 
Such a system can be described by
\begin{equation}
        \hat{H} = \frac{1}{2}\sum_i  \sigma^x_i - \Delta \sum_i n_i + \sum_{i \neq j} \frac{1}{((\textbf{r}_i - \textbf{r}_j)/R_b)^6}n_i n_j,
\end{equation}
where the local Hilbert spaces on the square lattice are spanned by $\{\ket{g}, \ket{r}\}$, where $\ket{g}$ denotes a local atom in the ground state, while $\ket{r}$ indicates it being in an excited (Rydberg-)state.  
We use the Rabi-frequency $\Omega$, the coupling constant of the first term, to fix the energy scale by setting it to unity together with the lattice distance.

This situation has recently been explored experimentally~\cite{ebadi2021_rydbergExperiment} as well as in numerical studies~\cite{Samajdar2020_Rydberg,ORourke2023_Rydberg_entanglement}. 
One of the central conclusions of one of the numerical studies has been the importance of the treatment of long-range tails of the Rydberg interactions~\cite{ORourke2023_Rydberg_entanglement}. This finding strongly suggests a reconsideration of studies on Rydberg atom arrays on various lattice geometries in with methods that can treat these long-range interactions faithfully. 

We investigate a slice of the phase diagram at fixed detuning $\Delta = 3$ whilst varying the Rydberg-blockade radius $R_b$.
In Fig.~\ref{fig:Rydbergs}, we show that several charge-density type phases, as well as disordered configurations, emerge. 
At small $R_b$ we find a checkerboard-state, that transitions into a striated phase ($R_b \approx 1.4-1.45$), in which the density of the checkerboard pattern is modulated. 
At $R_b = 1.5$ a homogeneous square phase emerges that eventually melts (via an intermediary phase) into a quantum disordered phase at $R_b = 1.9$. 
These results are consistent with the phase diagram obtained in~\cite{ORourke2023_Rydberg_entanglement}. 
In addition to the local Rydberg densities, we show the expectation values of the energy for the different ground states of the Rydberg atom array model. 
As expected, the slope of the energy density becomes flatter as the overall density of Rydberg excitations becomes smaller at a larger blockade radius $R_b$.

These results demonstrate that sampling-based finite PEPS calculation can be used comparatively easily to study particular models of long-range interacting Rydberg atom arrays.
This makes it possible to investigate the phase diagram for these models on other lattice geometries~\cite{samajdar2021_ryd_kagome} and aspect ratios as well as to study the influence of finite-size effects on the stabilization of phases. Particularly interesting for these investigations are the situations in which the Rydberg models are proposed to host topologically ordered ground states~\cite{Verresen21_topoRyd, Semenghini21_topoRyd_Exp}.

\section{conclusions and discussion}
In this work, we have discussed a framework for making use of sampling methods for finite PEPS calculations. 
To this end, we have pointed out how to solve the equations for the optimization more efficiently using an approach from the neural network community called minSR~\cite{chen2024minSR}.
We further have pointed out that variational bounds can be obtained using the finite PEPS in the sampling approach - which we then proceeded to use to verify that the standard approximations made are highly accurate. 
We then moved on to discuss complexity issues that might arise in this approach and have given a tentative quantity of the physical PEPS, that is correlated with the contraction complexity of its samples and have introduced a new initialization strategy based on these considerations.
Finally, we applied the finite PEPS in the context of a chiral spin liquid and of long-range interacting Rydberg atom arrays. We have pointed out that certain long-range interactions can be treated trivially in the sampling PEPS approach.

This demonstration on Rydberg atom arrays opens the door to the accurate numerical study of a multitude of related situations from frustrated geometries and topological order in Rydberg atoms~\cite{Verresen21_topoRyd, Semenghini21_topoRyd_Exp}, to phases of dipolar gases in cold atom platforms~\cite{Lahaye_2009_dipolar_gases}, and Nitrogen-vacancy centers~\cite{NV_centers1,NV_centers2}. 
One crucial application of this is the generation of digital twins of state-of-the-art quantum simulation platforms, which plays an important role in benchmarking these powerful experiments.

We note that a technique recently proposed in the context of iPEPS, for treating \textit{bra}- and \textit{ket}-layer separately~\cite{naumann2025variationallyoptimizinginfiniteprojected} might be useful to reduce computational cost in the double-layer boundary-MPS during the direct sampling procedure.

An important work for the future is an extensive comparison between the different methods, like different flavours of PEPS, NQS \cite{Carleo17_NQS}, TTNs \cite{Shi06_TTN,Tagliacozzo09_TTN, Rizzi19_anthology}, augmented TTNs \cite{augm_trees}, etc., for ground states calculations of two dimensional quantum systems.

Another area worth exploring further is how the single-layer contractability relates to the entanglement in the state represented by the finite PEPS. 
We suspect that given an entanglement entropy $S$, there exists a finite PEPS that will minimize the hardness of contracting its samples.

\paragraph*{Algorithm and open source code.}
An implementation of the algorithms discussed here is available as open source libraries ~\cite{quantumNaturalfPEPS, quantumNaturalgradient}.
Parts of them are making use of the ITensor library \cite{itensor, itensor-r0.3}.
All scripts used to generate the data for this analysis, along with the resulting simulation data, are available on Zenodo~\cite{peps2025zenodo}.

\section{Acknoledgements}

The authors thank Niklas Tausendpfund for generating the tree-tensor-network data used for comparison with the package TTN.jl \cite{TTN.jl}. 

This work was partially funded by the Deutsche Forschungsgemeinschaft (DFG, German Research Foundation) via Project-ID 277101999 -- CRC network TRR 183 (``Entangled states of matter'') and under Germany’s Excellence Strategy – Cluster of Excellence Matter and Light for Quantum Computing (ML4Q) EXC 2004/1 – 390534769,
and by the Horizon Europe programme HORIZON-CL4-2023-DIGITAL-EMERGING-01-CNECT via the project 101135699  (SPINUS) and Horizon Europe programme HORIZON-CL4-2022-QUANTUM-02-SGA via the project 101113690 (PASQuanS2.1).
E. L. W. thanks the Studienstiftung des deutschen Volkes for support. 
D.A. acknowledge funding by the German Federal Ministry of Education and Research (BMBF) for support under the thematic programme ``Quantum technologies -- from the basics to the market'', project number 13N16202 ``Noise in Quantum Algorithms (NiQ)''.

The authors gratefully acknowledge the Gauss Centre for Supercomputing e.V. (www.gauss-centre.eu) for funding this project by providing computing time through the John von Neumann Institute for Computing (NIC) on the GCS Supercomputer JUWELS~\cite{JUWELS} (Grant NeTeNeSyQuMa) and the FZ Jülich for computing time on JURECA~\cite{JURECA2021} (institute project PGI-8) at Jülich Supercomputing Centre (JSC).

\bibliography{literature.bib}
\newpage
\onecolumngrid
\appendix

\section{Sampling-based PEPS schemes can yield a variational upper bound on the energy}

To further investigate the effect of different approximations of the wave function amplitude $\Psi(\textbf{S})$ on the variational upper bound of the ground state energy, we compare the energy estimates obtained using two different computational approaches. As described in the main text, the faster method reuses the environments computed during boundary-MPS contraction to evaluate the local estimator $E_\text{loc}(\mathbf{S} )$, whereas the slower, variational approach enforces a consistent wave function approximation, ensuring a strict upper bound for the ground state energy.

\label{app:true_variational}
\begin{figure*}[ht]
    \subfloat[]{\includegraphics[width=0.45\textwidth]{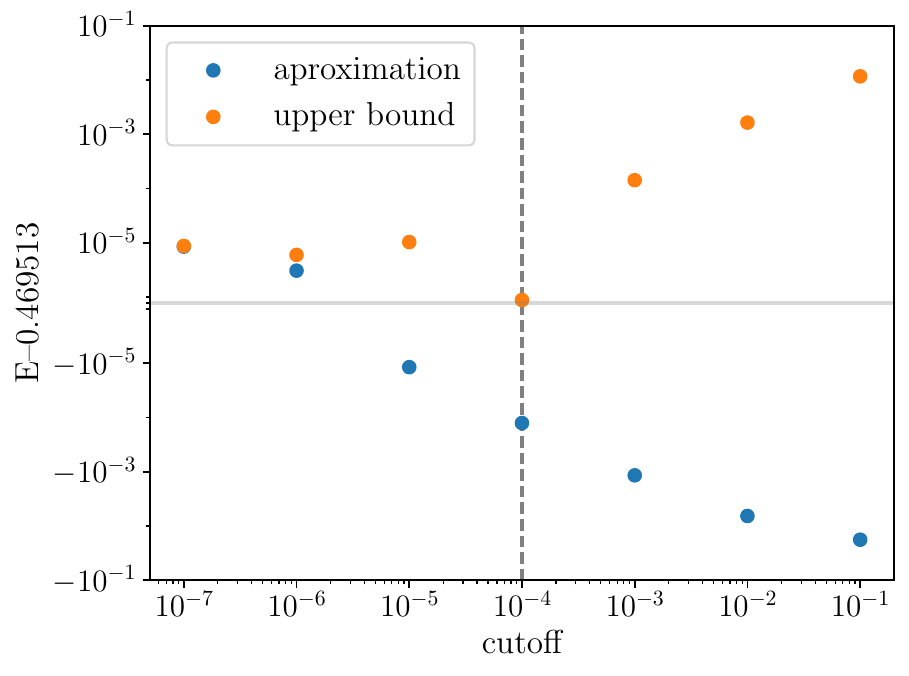}\label{fig:true_variational}}
    \subfloat[]{\includegraphics[width=0.45\textwidth]{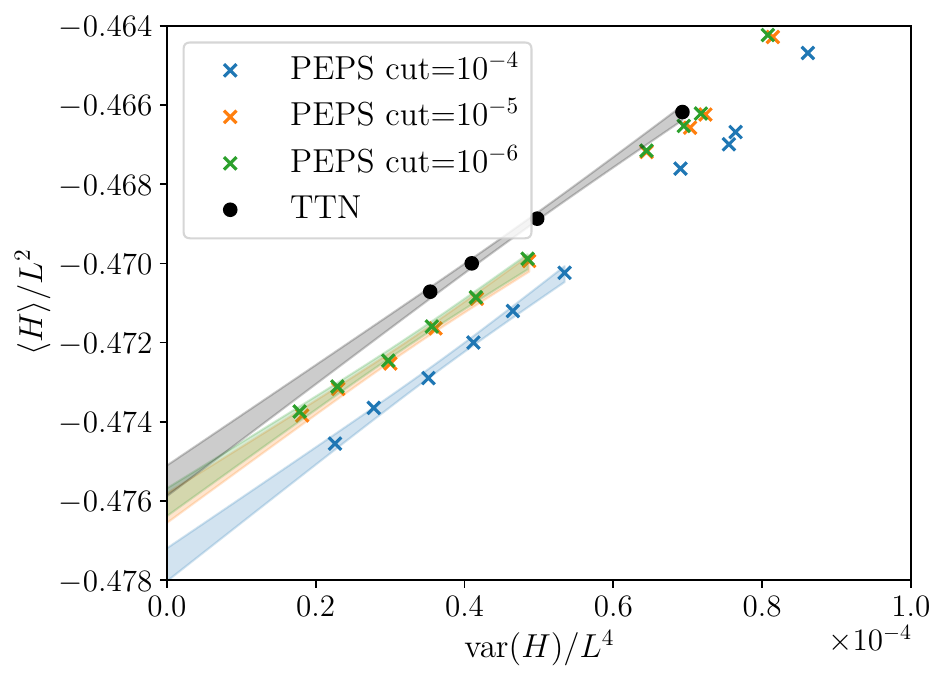} \label{fig:J1J2_optims:trees_vs_fpeps:cut}}
    
    \caption{(a) For a set of $10^4$ samples at $L=10$ the energy was computed for different cutoffs in two different ways: (blue) the faster way, which takes advantage of the environments to compute the $\Psi(\mathbf{S})$ needed to compute $E_\text{loc}(\mathbf{S} )$, and (orange) the slower way which always uses the same environments to compute an upper bound for the ground state energy. Note that the same samples were used for all calculations. The lowest variational energy value was used as the offset for the plot. (b) The energy vs variance plot for $L=16$ is displayed in Fig.~\ref{fig:J1J2_optims:trees_vs_fpeps} but calculated to a different accuracy using the different cutoffs for $10^5$ samples using the fast approximate method.} 
    \label{fig:true_variational2}
\end{figure*}%

Figure~\ref{fig:true_variational} presents the results for a set of $10^4$ samples, showing the energy deviation from the lowest obtained variational energy as a function of the contraction cutoff. For very small cutoffs ($10^{-7}$ to $10^{-5}$), both methods yield nearly identical energy estimates, suggesting that the approximation error is negligible in this regime. As the cutoff increases, deviations increase polynomially in the cutoff, with the fast method producing slightly lower energy estimates than the upper-bound method. This indicates that the ansatz indeed exploits inconsistencies in the wave function approximations to achieve artificially lower energy values.

Notably, at a cutoff of $10^{-4}$, the same value used during optimization, the strict variational calculation achieves its lowest value. Since $\braket{S|\Psi(\text{cutoff})}=E^u[i]( \text{cutoff}) \cdot E^l[i+1](\text{cutoff})$ was indirectly optimized using the cut environments $E$ it makes sense that its energy would perform best.

These results highlight the importance of choosing an appropriate contraction cutoff: while excessively tight cutoffs increase computational cost without significant accuracy gains, too loose a cutoff compromises the variational nature of the ansatz. The optimal choice depends on balancing these factors to ensure efficient yet reliable energy estimates.

An alternative approach for estimating an upper bound to the ground state energy involves a different, unbiased estimator that is computationally less demanding. Specifically, the method proposed at the end of Sec.~\ref{sec:reusing_envs} can be utilized. In this approach, the expectation values of the Hamiltonian terms are evaluated on a rotated basis where the Hamiltonian terms become diagonal. As a result, the computation of $E_\text{loc}(\mathbf{S})$ becomes independent of the wave function, eliminating the need for different approximations of $\Psi(S)_i$ and the errors that come with them.

This estimator is unbiased because the finite PEPS was not optimized using this method, ensuring that the optimization process could not exploit discrepancies between the different wave functions. The unbiased estimator yields $\braket{H}=-187.907 \pm 0.042$, which is in agreement with the other estimates. This consistency suggests that the approximations employed are sufficiently accurate to prevent significant deviations between different approximation schemes. Consequently, the use of additional methods appears unnecessary unless there is a reason to question the validity of the primary method.

More practical than generating expensive variational upper bounds is to vary the cutoff and see if energy and variance converge. For example, the PEPS states presented in Fig.~\ref{fig:J1J2_optims:trees_vs_fpeps:cut} were optimized using a cutoff value of $10^{-4}$. However, evaluating the energies obtained with this cutoff reveals discrepancies with the Tree Tensor Network (TTN) results, indicating inaccurate ground-state energies. This discrepancy disappears once the accuracy is increased. This observation underscores the necessity for meticulous care in PEPS calculations. In particular, ensuring sufficient accuracy in environment approximations is crucial to prevent the ansatz from exploiting numerical errors introduced during truncation.

\section{Direct sampling of PEPS}
\label{app:direct sampling of PEPS}

The direct-sampling scheme generates a many-body configuration $\textbf{S}$ with a probability $p(\textbf{S})$, which serves as an approximation of $p_{\Psi}(\textbf{S}) = \frac{|\Psi(\textbf{S})|^2}{\braket{\Psi|\Psi}}$. The discrepancy between $p(\textbf{S})$ and $p_{\Psi}(\textbf{S})$ can be corrected using importance sampling, as will be discussed later.

To proceed, we introduce the shorthand notation $\textbf{S}_i := \{s_{i,1}, \dots, s_{i,L_y} \}$ to represent the collection of local configurations on the $i$-th row. This allows us to express the probability of a configuration as
\begin{equation}
    p(\textbf{S}) = p(\textbf{S}_1) \prod_{i=1}^{L_y-1} p(\textbf{S}_{i+1} | \textbf{S}_{<i+1}),
    \label{eq:prob_from_cond_prob}
\end{equation}
where $p(\textbf{S}_i | \textbf{S}_{<i})$ represents the conditional probability given the configurations on the uppermost $i-1$ rows, denoted by $\textbf{S}_{<i}= \{ \textbf{S}_1, \dots, \textbf{S}_{i-1} \}$.

We generate each of the probabilities in Eq.~\eqref{eq:prob_from_cond_prob} sequentially, beginning with $p(\textbf{S}_1)$. This probability corresponds to the reduced density matrix of the topmost row of the PEPS, denoted as $\rho[1]$. As this step involves tracing out all degrees of freedom except those on the first row, an approximate contraction of the double-layer tensor PEPS network for all rows except the first is required.

\begin{figure}[ht]
    \includegraphics[width=0.49\textwidth]{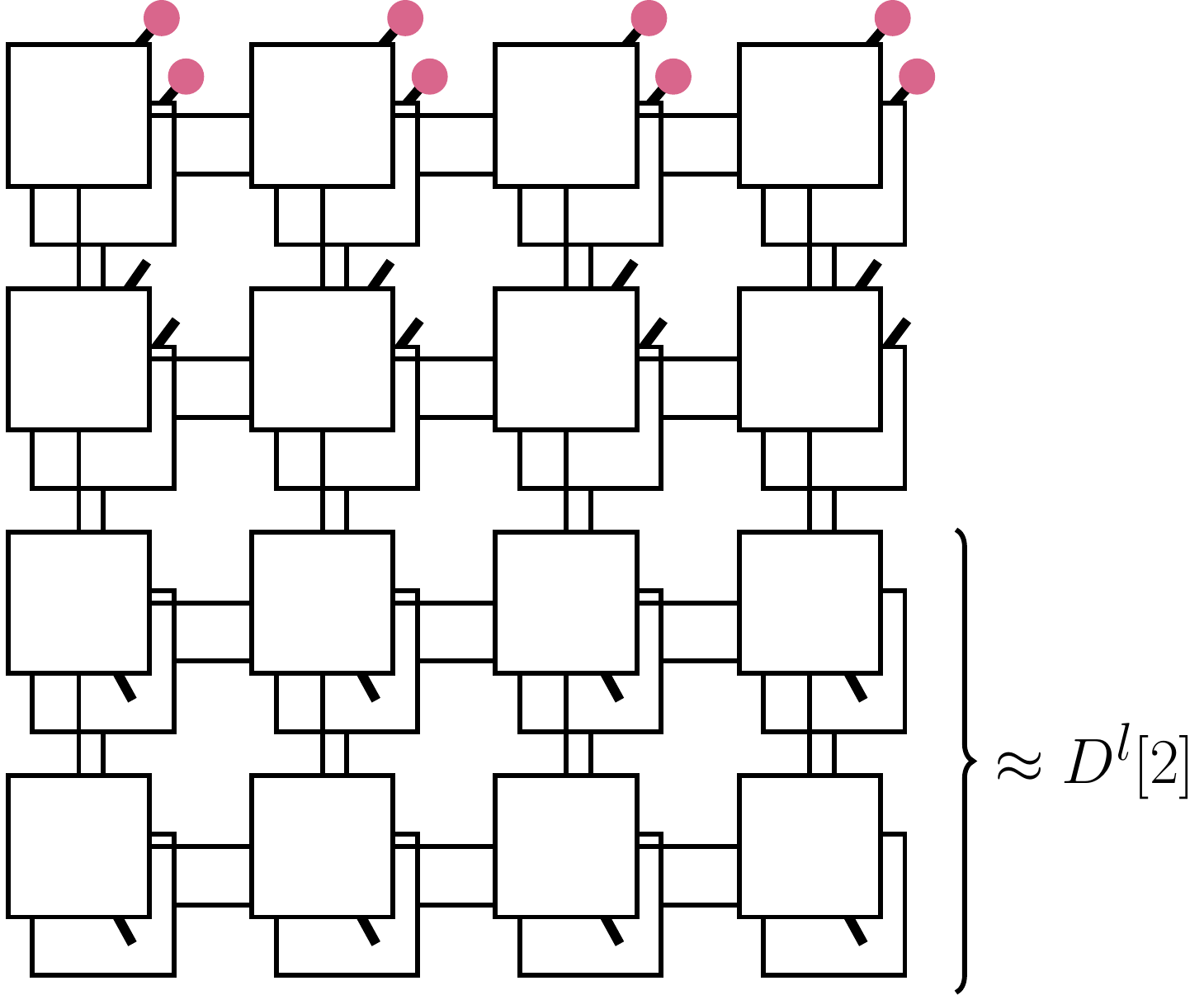}
    \caption{Illustration of the conditional reduced density matrix for the second row $\rho_{\textbf{S}_1}[2]$ used in the direct sampling scheme. The spin configurations of the first row are fixed to configurations $\textbf{S}_1$, which is illustrated by the colored dots. The physical legs of the second row are left open (on the support of $\rho_{\textbf{S}_1}[2]$), while all physical legs on the lower rows are traced out. The contraction of the two lowermost rows, on which the physical indices are traced out, are in our calculation approximated by a boundary-MPS $D^l[2]$.}
    \label{fig:sampling}
\end{figure}

To achieve this, we define $D^l[i]$ as the double-layer contraction of the $i$ lowermost rows, computed using the boundary-MPS method described above (cf. Fig.~\ref{fig:sampling}). In practice, we truncate the bond dimension of $D^l[i]$ to $D_c^{\text{double}}$. Additional comments on this double-layer boundary can be found at the end of this section.

An approximation of the reduced density matrix for the first row can thus be expressed as
\begin{equation}
\begin{split}
    \rho[1] &= T^u[1] \cdot D^d[2] \cdot (T^u)^*[1] \\
            &= \begin{gathered}
        \includegraphics[height=6.0cm]{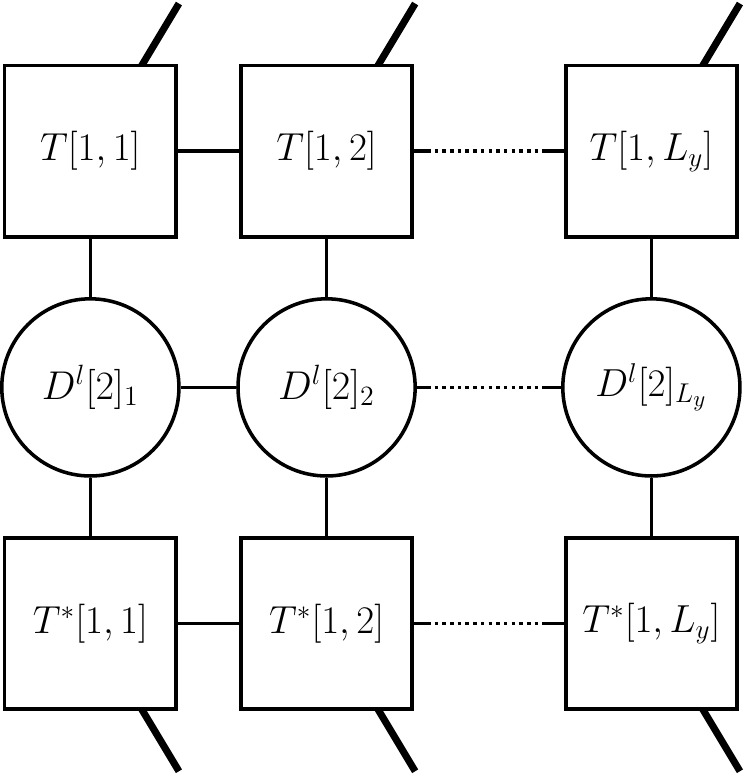}
    \end{gathered}, 
\end{split}
\label{eq:rho}
\end{equation}
where $T^u[i]$ represents the product of all local PEPS tensors in the $i$-th uppermost row, as illustrated in the equation above. 

It is noteworthy that this reduced-density matrix possesses a one-dimensional structure. Consequently, efficient methods for sampling MPS wavefunctions can be directly applied to generate samples $\textbf{S}_1$ from the probability distribution $p(\textbf{S}_1)$~\cite{Vidal12_sampling}. This established MPS technique resembles the direct-sampling approach for PEPS discussed here, as the probability $p(\textbf{S}_1)$ is similarly expressed as a product of conditional probabilities, analogous to Eq.~\eqref{eq:prob_from_cond_prob}, which are evaluated sequentially from left to right (or vice versa).


We now proceed to the generation of the conditional probabilities $p(\textbf{S}_{i}|\textbf{S}_{<i})$, which are derived from the conditional reduced density matrix $\rho_{\textbf{S}_{<i}}[i]$ of the $i$-th row, as illustrated in Fig.~\ref{fig:sampling}. The support of $\rho_{\textbf{S}_{<i}}[i]$ consists of the degrees of freedom on the $i$-th row, while all degrees of freedom on the rows above are fixed to the configurations $\textbf{S}_{<i}$. 

To construct this conditional reduced density matrix, we combine the $i$ uppermost rows using the boundary-MPS method, resulting in
\begin{equation}
\begin{split}
    T^u_{\textbf{S}_{< i}}[i] :&= \begin{gathered}
         \includegraphics[height=2.2cm]{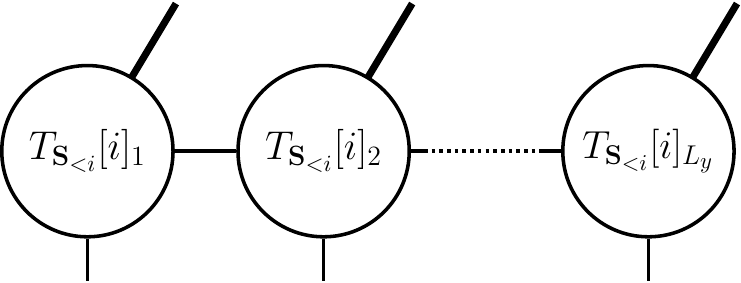}
     \end{gathered}\\
     &\approx \begin{gathered}
         \includegraphics[height=6.5cm]{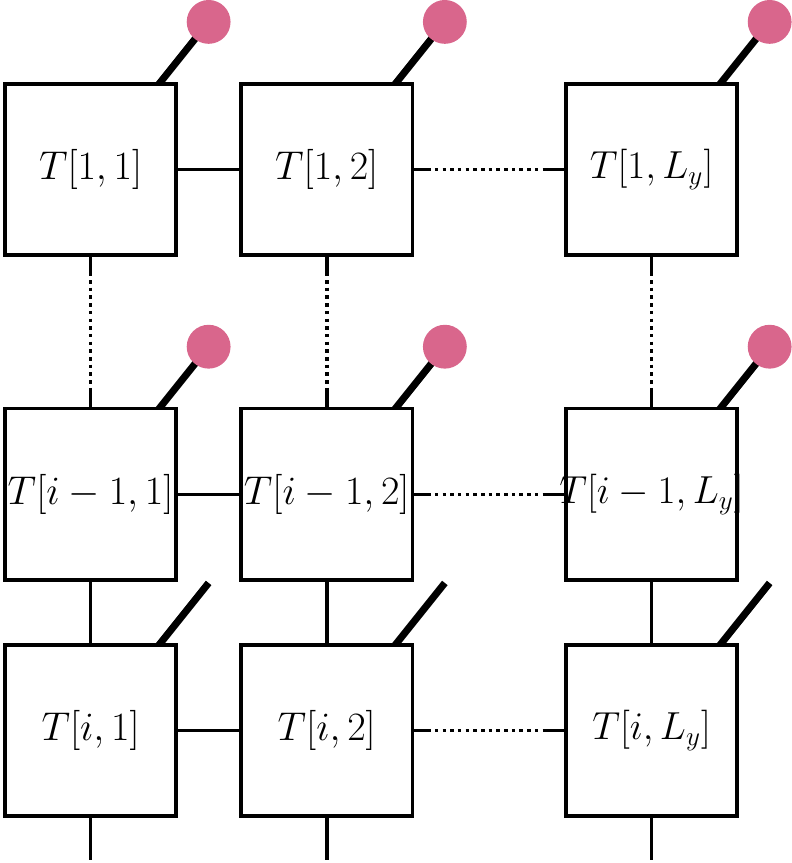}
     \end{gathered},
\end{split}
\end{equation}
where the bond dimension is truncated to $D_s$ to ensure computational efficiency. Note that in practice, $D_s$ can be chosen to be significantly smaller than $D_c$. All truncations are performed using the algorithm described in \cite{McCulloch_2007_DMRG_MPSMPO}.

All rows below the $i$-th one are traced out, which is represented using the double-layer boundary-MPS $D^l[L_x-i]$. This approach allows for an efficient and stable computation of the conditional probabilities required for the sampling procedure.
With these objects, we can express the conditional reduced density matrix $\rho_{\textbf{S}_{<i}}[i]$ as
\begin{equation}
    \rho_{\textbf{S}_{<i}}[i] = T^u_{\textbf{S}_{<i}}[i] \cdot D^l[i+1]\cdot (T^u_{\textbf{S}_{<i}})^*[i],
\end{equation}
which has the same form as Eq.~\eqref{eq:rho} such that we can again use the sampling algorithms for MPS to obtain the conditional probability  $p(\textbf{S}_i | \textbf{S}_{<i})$ from it. This is summarized in $Alg.~\ref{alg:sampling}$.

\begin{algorithm}[H]
\caption{Sampling algorithm}
\label{alg:sampling}
\begin{algorithmic}
\Function {sample}{$T, D^l; D_c, D_s$}
    \State $E^u[0] \gets 1$
    \For{$i = 1, \dots, L$}
        \State $T^u_{\textbf{S}_{<i}}[i] \gets \text{mul}(T[i,:],  E^u[i-1]; D_s)$
        \State $S_i \gets \text{sample}(T^u_{\textbf{S}_{<i}}[i], D^l[i+1])$
        \State $T_\text{proj}[i] \gets \text{proj}(T[i,:], S_i)$
        \State $E^u[i] \gets \text{mul}(E^u[i-1], T_\text{proj}[i]; D_c)$
    \EndFor
    \State \Return $\mathbf{S}, E^u$
\EndFunction
\end{algorithmic}
\end{algorithm}

We close this summary of the direct sampling procedure with a few comments. Firstly, the direct sampling procedure involves the calculation of boundary-MPS approximation for double layers of PEPS.
However, as was noted in~\cite{Vieijra21_directsampling_PEPS}, we can get away with taking small values for the environment bond dimension $D_c^{\text{double}} \sim D$ of this double-layer boundary-MPS.
Additionally, these double-layer environments have to be calculated only once and can then be reused to generate an arbitrary number of samples for the corresponding PEPS.
Due to this fact, the calculation of the double-layer boundary-MPS for small enough $D$ only accounts for a small fraction of the computational time used.

Secondly, as mentioned at the beginning of this section, the fact that we obtain any sample $\textbf{S}$ with a probability $p(\textbf{S})$ which is an approximation of the actual probability $p_\Psi(\textbf{S})$ can be corrected for with an additional factor of $\frac{p_\Psi(\textbf{S})}{p(\textbf{S})}$ in Eq.~\eqref{eq:local_estimator}. Note that once we have a sample $\textbf{S}$ we can obtain $\Psi(\textbf{S})$ needed for $p_\Psi(\textbf{S})$ accurately with a single layer contraction.

The computation of double-layer environments becomes increasingly demanding for larger bond bond dimensions, taking up a sizable share of the computational time for $D\ge7$.
This issue can be mitigated by computing the double-layer environments asynchronously. This approach generates samples using previously computed double-layer environments while updated environments are simultaneously computed. Although asynchronous generation introduces slight inaccuracies due to outdated environments, these deviations are corrected through importance sampling. Moreover, the magnitude of these errors can be continuously monitored by evaluating the statistical error of the energy, ensuring that statistical biases remain within acceptable limits. In practice for a $L=16$ $D=8$ PEPS with $2000$ samples, the double layer environments will lag 5 optimization steps behind while not causing any significant change in the error metrics.

\section{Real-Time Evolution}
\label{app:real_time_evo}
The equations introduced in Sec.~\ref{sec:imaginary_schrodinger_eq} can, in principle, be applied to real-time evolution. However, two major challenges prevent achieving high-fidelity results in practice.

The first challenge arises from the typically linear growth of entanglement entropy during real-time evolution. PEPS inherently obey an area-law scaling of entanglement entropy and are limited to modest bond dimensions, generally ranging from $1$ to $10$, due to computational constraints. Consequently, accurate simulations over long time intervals are usually infeasible. Exceptions occur in special cases where entanglement entropy grows sublinearly, such as the domain-wall dynamics recently examined in Ref.~\cite{krinitsin2024_rougheningdynamicsinterfacestwodimensional}.

The second challenge relates to the number of samples required for accurately solving the TDVP equations. In optimization scenarios, the fidelity of individual evolution steps is less critical, provided convergence to the correct state is eventually achieved. In contrast, real-time evolution is highly sensitive to fidelity losses, which accumulate exponentially over multiple steps. Ensuring high-fidelity evolution steps thus requires a substantial number of samples, often several orders of magnitude larger than the number of variational parameters. Schmidt \etal~\cite{schmitt2020quantum}, for instance, used around $10^6$ samples for an ansatz with roughly $10^3$ parameters. A PEPS with moderate system size ($L = 10$) and bond dimension ($D = 6$) contains substantially more parameters ($N_p\approx10^5$). Therefore, meeting the sampling requirements in practical scenarios is expected to be a significant computational challenge.

\section{Efficient computation of geometric entanglement}
Several methods exist to compute the geometric entanglement for finite PEPS. In this study, a two-step approach is employed. The first step provides an approximate product state that maximizes the overlap
\begin{align}
\Lambda_{\max} &= \max_{\ket{\phi}} \left| \bra{\phi} \Psi \rangle \right|^2, \nonumber\\
S_G (\ket{\psi}) &= -\frac{1}{N}\log_2 \Lambda_{\max}\;,
\end{align}
while the second step refines this initial approximation.

To efficiently approximate the optimal state $\ket{\phi} = \bigotimes_{i=1}^{L^2} \ket{\phi_i}$, a sampling-based strategy is utilized. Rather than sampling directly from $\ket{\Psi}$ in the $z$-basis, the same algorithm is used, but now the spins are sequentially optimized to maximize the overlap with the conditional reduced density matrix. This selection is performed iteratively by minimizing
\begin{align}
\min_{\ket{\phi_j}}\bra{\Psi} (\bigotimes_{i=1}^{j}\ket{\phi_i}\bra{\phi_i})\ket{\Psi}\;.
\end{align}
After obtaining this initial approximation, the overlap $\left| \bra{\phi} \Psi \rangle \right|^2$ can be further optimized through a sweep across all $\ket{\phi_i}$. During this sweep, each $\ket{\phi_i}$ is individually optimized by maximizing its overlap, allowing for efficient reuse of previously computed environments.

\label{app:geom_ent}
\end{document}